%% file: main.tex
\documentclass[format=acmlarge,11pt,review=false, authorversion,nonacm]{acmart}
\usepackage{acm-ec-21}
\usepackage{booktabs} 
\usepackage[ruled]{algorithm2e} 
\usepackage{thmtools, thm-restate}

\SetAlFnt{\small}
\SetAlCapFnt{\small}
\SetAlCapNameFnt{\small}
\SetAlCapHSkip{0pt}
\IncMargin{-\parindent}
\DeclareMathOperator{\diag}{diag} 
\DeclareMathOperator{\sgn}{sgn} 
\setcitestyle{acmnumeric}

\newtheorem{conj}{Conjecture}[section]

\makeatletter
\let\@authorsaddresses\@empty
\makeatother

\setcopyright{none}
\usepackage{caption}
\usepackage{subfigure}
\usepackage{multirow}
\usepackage{graphicx} 

\title{Co-evolution of Opinion and Social Tie Dynamics Towards Structural Balance}

\author{Haotian Wang}
\affiliation{
  \institution{Rutgers University, hw487@cs.rutgers.edu}
}
\author{Feng Luo}
\affiliation{
  \institution{Rutgers University, fluo@math.rutgers.edu}
}
\author{Jie Gao}
\affiliation{
  \institution{Rutgers University, jg1555@rutgers.edu}
}
\begin{abstract}
In this paper, we propose co-evolution models for both dynamics of opinions (people's view on a particular topic) and dynamics of social appraisals (the approval or disapproval towards each other). Opinion dynamics and dynamics of signed networks, respectively, have been extensively studied. 
We propose a co-evolution model, where each vertex $i$ in the network has a current opinion vector $v_i$ and each edge $(i, j)$ has a weight $w_{ij}$ that models the relationship between $i, j$. The system evolves as opinions and edge weights are updated over time by the following rules: 
\begin{itemize}
    \item \emph{Opinion dynamics:} The opinion of agent $i$ is updated as a linear combination of its current opinion and the weighted sum of neighbors' opinions with coefficients in matrix $W=[w_{ij}]$. 
    \item \emph{Appraisal dynamics:} The appraisal $w_{ij}$ is updated as a linear combination of its current value and the agreement of the opinions of agents $i$ and $j$. The agreement of opinion $v_i$ and $v_j$ is taken as the dot product $v_i 
    \cdot v_j$.
\end{itemize} 
We are interested in characterizing the long-time behavior of the dynamic model -- i.e., whether edge weights evolve to have stable signs (positive or negative) and structural balance (the multiplication of weights on any triangle is non-negative).

Our main theoretical result solves the above dynamic system with time-evolving opinions $V(t)=[v_1(t), \cdots, v_n(t)]$ and social tie weights $W(t)=[w_{ij}(t)]_{n\times n}$. For a generic initial opinion vector $V(0)$ and weight matrix $W(0)$, one of the two phenomena must occur at the limit. The first one is that both sign stability and structural balance (for any triangle with individual $i, j, k$,  $w_{ij}w_{jk}w_{ki}\geq 0$) occur. In the special case that $V(0)$ is an eigenvector of $W(0)$, we are able to obtain the explicit solution to the co-evolution equation and give exact estimates on the blowup time and rate convergence. The second one is that all the opinions converge to $0$, i.e., $\lim_{t\rightarrow \infty}|V(t)|=0$. 

We also performed extensive simulations to examine how different initial conditions affect the network evolution.  Of particular interest is that our dynamic model can be used to faithfully detect community structures. On real-world graphs, with a small number of seeds initially assigned ground truth opinions, the dynamic model successfully discovers the final community structure. The model sheds lights on why community structure emerges and becomes a widely observed, sustainable property in complex networks. 
\end{abstract}

\begin{document}

\begin{titlepage}

\maketitle

\end{titlepage}

\section{Introduction}

We live in a continuously changing world in which social interactions dynamically shape who we are, how we view the world and what decisions to make. Various social processes, naturally intertwined, operate on both the properties of individuals and social ties among them. Social influence, for example, describes how people's behaviors, habits or opinions are shifted by those of their neighbors. 
Social influence leads to homophily ( similarity of node attributes between friends) and leaves traces in the network structure such as high clustering coefficient and triadic closure (there are likely social ties among one's friends). 

Social influence and homophily, however, do not fully interpret the global network structure. One of the widely observed structural properties in social networks is the community structure. Nodes within the same community are densely connected and nodes from different communities are sparsely connected. Community detection is an important topic in social network analysis and has been investigated extensively~\cite{Pares2017-yq,Yang2016-gn,Newman2004-og,Newman2004-js,Newman2004-jn,Guimera2004-vm,Zhang2014-vs,Von_Luxburg2007-wb,Newman2006-sa,Hastings2006-sn,Karrer2011-ms}. But why does community structure emerge and become a persistent feature? Are there social processes that encourage or maintain the community structure? 

In the literature, there have been a lot of studies of opinion dynamics, where people's behaviors, habits or opinions are influenced by those of their neighbors. Network models that capture social influence, such as French-DeGroot model~\cite{French1956-bk, Antal2005-lu},  Friedkin-Johnsen model~\cite
{Friedkin1990-wl}, 
Ku$\l$akowski et al. model~\cite{Kulakowski2005-is, Marvel2011-we}  naturally converge to global consensus. Actually it seems to be non-trivial to create non-homogeneous outcomes in these models, while in reality social groups often fail to reach consensus and exhibit clustering of opinions and other irregular behaviors. Getting a model that may produce community cleavage~\cite{Friedkin2015-qc} or diversity~\cite{Kurahashi-Nakamura2016-ys} often requires specifically engineered or planted elements in a rather explicit manner. For example, bounded confidence models~\cite{Mas2010-za,Krause2000-no,Hegselmann2002-ud,Deffuant2000-pb} 
limit social influence only within pairs with opinions sufficiently close. 
Other models introduce stubborn nodes whose opinions remain unchanged throughout the process.

\begin{figure}[htbp]
	\centering
\includegraphics[width=0.5\columnwidth]{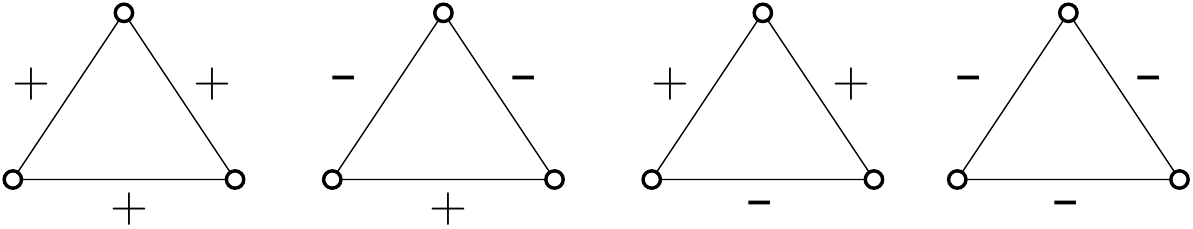}\vspace*{-3mm}
\caption{Structural balance theory: the first two triangles are stable while the last two are not.}\label{fig:structural-balance}
\end{figure}

The extreme case of community structure is dictated by structural balance~\cite{Heider1946-bt, Heider1982-rg,Cartwright1956-fc, davis1967clustering}, a common phenomenon observed in many social relationships. 
The notion was first introduced in a seminal paper by Heider in the 1940s in social psychology.   It describes the stability of human relations among three individuals when there are only two types of social ties: positive ties describe friendship (or sharing common opinions) and negative ties describe hostility (or having opposite opinions). 
Heider's axioms state that among three individuals, only two kinds of triangles are stable: the triangles where all three ties are positive, indicating all three individuals are mutual friends; and the triangle with two negative edges and one positive edge, describing the folklore that ``the enemy of your enemy is your friend.''
The other two types of triangles (e.g., a triangle of one negative tie and two positive ties, or three mutually hostile individuals) incur emotional stress or are not strategically optimal. See Figure~\ref{fig:structural-balance}.  
Thus they are not socially stable -- over time, they break and change to the stable ones. In fact, the structural balance theory not only describes the local property in a signed complete network, but also predicts the global network behavior -- the only type of network in which all triangles are stable must have the nodes partitioned into two camps, inside each of which the edges are all positive and between them the edges are negative (one of the camps can be empty).

The structural balance theory only describes the equilibrium state and does not provide any model on evolution or dynamics -- what happens when a network has unstable triangles? Follow up work on structural balance dynamic models~\cite{Antal2005-lu,Marvel2011-we, Kulakowski2005-is,cisnerosvelarde2019structural, teixeira2017emergence} update the sign/weight of an edge  $(i, j)$ towards a more balanced triangle by considering the sign/weights of neighboring edges $(i, k)$ and $(k, j)$ in a triangle $\triangle i j k$. 

These two threads of research, opinion dynamics and structural balance dynamics,
are currently orthogonal to each other. 
In opinion dynamics, opinions on vertices are influenced by each other through edges but researchers struggle to maneuver the model to create community cleavage. In structural balance dynamics, edge weights are dynamically updated to meet structural balance, which is explicitly coded as the projected outcome and optimization objective. Holme and Newman~\cite{holme2006nonequilibrium} presented a simple model of this combination without any theoretical analysis. In this paper, we propose co-evolution models for both dynamics of opinions (people's view on a variety of topics) and dynamics of social appraisals (the approval or disapproval towards each other). We show that by using two simple rules, node opinions evolve into opposing communities and structural balance naturally emerges.



\subsection{Our Contribution}

In this paper, we consider the co-evolution of opinions via social influence and tie strength/appraisal/sign updates by discrepancies of node opinions. We assume a set of $n$ individuals where individual $i$ has an opinion $v_i \in \mathbb{R} $, and an appraisal matrix $W=[w_{ij}]$, where $w_{ij}$ is interpreted as the influence from individual $j$ on individual $i$. Here $w_{ij}$ does not need to be non-negative and takes values in $\mathbb{R}$. We consider two update rules: 
\begin{itemize}
    \item \emph{Opinion dynamics:} The opinion of $i$ is updated as a linear combination of its current opinion and the weighted sum of neighbors' opinions with coefficients in matrix $W$. 
    \item \emph{Appraisal dynamics:} The appraisal $w_{ij}$ is updated as a linear combination of its current value and the agreement of the opinion of $i$ and $j$. The agreement of opinion $v_i$ and $v_j$ is taken as the dot product $v_i 
    \cdot v_j$.
\end{itemize}
The opinion dynamics model is similar to classical social influence models such as the DeGroot model~\cite{French1956-bk} and Friedkin-Johnsen model~\cite{Friedkin1990-wl}, except that the edge weight matrix $W$ is dynamic as well. The model for appraisal dynamics is motivated by tie dynamics that can be traced back to Schelling's model of residential segregation~\cite{Schelling1971-eb}. In modern society, tie changes on Facebook~\cite{Sibona2014-ni} and Twitter~\cite{Xu:2013:SBT:2441776.2441875,kivran2011impact} can be easily triggered by disparities on their  opinions~\cite{John2015-az,Sibona2014-ni}, especially among the users who are most politically engaged.

Our goal is to analyze the evolution and in particular, the conditions that lead to \emph{sign stability} and \emph{structural balance} -- for any triangle with individual $i, j, k$,  $w_{ij}w_{jk}w_{ki}\geq 0$. We call a network to reach strictly structural balance if  $w_{ij}w_{jk}w_{ki}> 0$, $\forall i, j, k$. 
A structurally balanced network has two possible states: \emph{harmony}, when all edges are positive; and \emph{polarization}, when there are two communities with only positive ties within each community and negative ties across the two communities. 

We show that with our dynamics model, $W(t)$ evolves by the following matrix Riccati Equation~\cite{Abou-Kandil2003-um} $W' = WW^T + C$, where $C= V(0)V(0)^T - W(0)W(0)^T$ is a symmetric constant $n \times n$ matrix if $W(0)$ is. Further the opinion vector $V(0)$ evolves by the differential equation $V''(t)=2|V|^2 \cdot V-C\cdot V$. 
Our main result is to analyze the asymptotic behavior of $V(t)$ and $W(t)$ and prove structural balance at the limit. Our results can be summarized in the following:
\begin{enumerate}
    \item By analyzing the evolving equation for $V(t)$, we show that either the network reaches strict structural balance or $|V(t)|\rightarrow 0$. To prove this limit behavior, one crucial observation is that the length of the opinion vector $|V(t)|^2$ is strictly convex, unless $V(t)\equiv 0$.
    \item We show how to solve the general matrix Riccati equation $W'=W^2+C$, $W(0)=B$, for any parameter $B, C$. In particular, the eigenvector corresponding to the largest eigenvalue of $W(t)$ encodes the two communities formed in the network; those with a positive value in the eigenvector versus those with a negative value in the eigenvector. As a byproduct of the analysis of $W(t)$, we also show that when $V(0)$ is an eigenvector of the initial matrix $W(0)$, $V(t)$ remains to be an eigenvector of $W(t)$ for and structural balance must occur in finite time. In this case we can write down exact evaluations on the blowup time and the rate convergence.
\end{enumerate}

The evolving of $W(t)$ by $W' = WW^T + C$ is strictly a generalization of the dynamic structural balance model by Marvel et al. ~\cite{Marvel2009-lx}. Their model captures the dynamics of edge appraisals by $W' = W^2$ and does not consider user opinions. The behavior of our model becomes much more complex, as the initial user opinions are factored into the system dynamics through the matrix $C$.


We also performed extensive simulations with different initial conditions and graph topology. We examined the network evolution on the final convergence state (harmony v.s., polarization) and convergence rate. We observed that a higher network density or a higher initial opinion magnitude, empirically, speeds up the convergence rate. 

We tested our dynamic model on two real world graphs (Karate club graph and a political blog graph~\cite{adamic2005political}). Both networks are known to have two communities with opposing opinions. A small number of seeds, randomly selected, are assigned with ground truth opinions and all other nodes start neural. The network evolution can successfully detect the final community structure and recover the ground truth with good accuracy. Apart from being a transparent and explainable label propagation algorithm, the model sheds lights on why community structure emerges and becomes a widely observed, sustainable property in complex networks. 

\section{Related Work}

\subsection{Opinion Dynamics and Social Influence} 

Opinions in a sociological viewpoint capture the cognitive orientation towards issues, events or other subjects, and mathematically represent signed attitudes or certainties of belief. Opinion dynamics is an extensively studied topic about how opinions change in a network setting with social influence from neighbors.
One of the first models of opinion dynamics, French-DeGroot model~\cite{French1956-bk, Antal2005-lu}, considers a discrete time process of opinion $\{v_1, v_2, \cdots, v_n\}$ for a group of $n$ individuals. An edge $(i, j)$ carries a non-negative weight $w_{ij}
\geq 0$. The opinion of node $i$ at time $t+1$ is updated by
$$v_i(t+1)=\sum_{j}w_{ij}v_j(t).$$
The weight matrix $W=[w_{ij}]$ is taken as a stochastic matrix. The dynamics can be written as $V(t+1)=WV(t)$.
The continuous-time counterpart is called the Abelson's model~\cite{Abelson1964-ny} where the dynamics is defined by 
\begin{equation}\label{ref:eqn-abelson}
    \frac{dV(t)}{dt}=-LV(t),
\end{equation}
where $V(t)=(v_1(t), v_2(t), \cdots v_n(t))^T$ and $L$ is the Laplacian matrix $L=I-W$.
Opinions following the French-DeGroot model or the Abelson's model typically converge unless the network is disconnected or there are stubborn nodes (with $w_{ii}=1$).

The most popular opinion dynamics model is probably the Friedkin-Johnsen model~\cite{Friedkin1990-wl}. It takes a stochastic matrix $W$ as the influence model, and a diagonal matrix $\Lambda=\diag(\lambda_1, \cdots, \lambda_n)$ where $\lambda_i\in [0, 1]$ is the susceptibility of individual $i$ to social influence. The opinions of the individuals are updated by the following process
$$V(t+1)=\Lambda W V(t)+(I-\Lambda)u,$$
where $u$ is a constant vector of the individuals' prejudices and is often taken as the initial opinion $V(0)$. When $\Lambda=I$ the model turns to French-DeGroot model. 

Most of the literature on these two models assume a fixed weight matrix $W$ and prove asymptotic convergence under favorable assumptions~\cite{Friedkin2011-et}. There have been extensions when $W$ is a time-varying matrix, but $W(t)$ is still independent of $V(t)$ (e.g.,~\cite{Blondel2005-uz}).

A significant deviation from the above family considers a time-varying matrix $W$, by incorporating the principle of homophily, that similar individuals interact more than dissimilar ones. 
This is called the bounded confidence model~\cite{Mas2010-za}. A few such models (Hegselmann-
Krause (HK) model~\cite{Krause2000-no,Hegselmann2002-ud}, Deffuant and Weisbuch~\cite{Deffuant2000-pb}) introduced a fixed range of confidence $d>0$: individual $i$ is insensitive to opinions that fall outside its confidence set $I_i=[v_i -d, v_i+d]$, and the opinion $v_i$ is only updated by the average opinion of those opinions within $I_i$. In other words, the matrix $W$ is derived from the set of opinions at time $t$ and thus co-evolves with the opinions. This model generates situations when the individuals converge to a set of different opinions, and has been extended to the multi-dimensional setting~\cite{Hegselmann2002-ud}.

All models above have only considered the case of positive influence, that the interactions of individuals change their opinions \emph{towards} each other. It has been argued in both social settings and many physical systems that there is negative or repulsive influence (repulsive interactions in biological systems~\cite{Coyte2015-ry} or collision avoidance in robot swarm formation~\cite{Romanczuk2012-sg}). Abelson~\cite{Abelson1967-nn} argued that any attempt to persuade a person may sometimes shift his or her opinion away from the persuader's opinion, called the boomerang effect~\cite{Allahverdyan2014-ja, Hovland1953-hc}. Bhawalkar et al.~\cite{bhawalkar2013coevolutionary} presented game-theoretic models of opinion formation in social networks by maximizing agreement with friends weighted by the strength of the relationships. Thus interactions between individuals with similar opinions move their opinions closer; interactions between individuals with opinions that are very different shift their opinions away from each other. Here the edge weights are fixed. Many models have included negative ties but they are still awaiting rigorous analysis~\cite{Salzarulo2006-ag, Baldassarri2007-zs, Macy2003-wq, Mark2003-it, he2018evolution}.
The most notable work in this direction is by Altafini~\cite{Altafini2013-kb, Altafini2012-lr}. The model starts to be similar to Abelson's model in Equation~(\ref{ref:eqn-abelson}) with a \emph{fixed} weight matrix $W$ except that the weights in $W$ do not need to be non-negative. The system is shown to be Lyapunov stable~\cite{Proskurnikov2016-hc} and studies have focused on the initial conditions of $W$ for the system to converge to harmony or polarization. The matrix $W$ is assumed to be either static or, in very recent studies~\cite{Proskurnikov2017-lr, Proskurnikov2016-hc,Hendrickx2014-qy}, time-varying (but independent of the opinion changes). 
The negative influence is closely related to signed networks and structural balance theory, which will be discussed next. 

\subsection{Structural Balance and Signed Networks}

Notice that the structural balance theory only describes the equilibrium state and does not provide any model on evolution or dynamics -- what happens when a network has unstable triangles? Follow up work proposed a few models, that can be categorized by discrete models or continuous models -- depending on whether the appraisal on a social tie takes binary values $\{+1, -1\}$ or a real number.
Antal et al.~\cite{Antal2005-lu} considered the discrete model where the sign of an edge is flipped if this produces more balanced triangles than unbalanced ones. The balanced graph is clearly a stable state but the dynamics also has many local optimals called jammed states~\cite{Antal2005-lu,Marvel2009-lx}. Andreida et al.~\cite{teixeira2017emergence} determine the sign of an edge according to the sign of the other two edges in the triangles to make more triangles balanced. Samin et al.~\cite{aref2020modeling} try to remove the minimum edges to make the graph balanced which is NP-hard problem.

In the continuous setting, the influence-based model~\cite{Marvel2011-we, Kulakowski2005-is} describes an influence process on a complete graph, in which an individual $i$ updates her appraisal of individual $j$ based on what others positively or negatively think of $j$. In other words, let us use $w_{ij}$ to describe the type of the social tie between two individuals $i, j$. $w_{ij}>0$ if $i, j$ are friends and $<0$ if they feel negative about each other. The absolute value of $w_{ij}$ describes the magnitude of the appraisal. The update rule says that the update to $w_{ij}$ will take value 
\begin{equation}\label{eq:dynamic-w}
\frac{d w_{ij}}{dt}=\sum_{k}w_{ik}\cdot w_{kj}.
\end{equation}
Specifically, when $w_{ik}$ and $w_{kj}$ have the same sign, the value of $w_{ij}$ is guided to the positive direction; when $w_{ik}$ and $w_{kj}$ have opposite signs, the value of $w_{ij}$ is guided to the negative direction. Both cases try to enforce a balanced triangle on $\{i, j, k\}$. Empirically, it has been observed that for essentially any initial value of $W$, as the matrix where the $(i, j)$ element is $w_{ij}$, the system reached a balanced pattern in finite time. In~\cite{Marvel2011-we}, Marvel et al. proved that for a random initial matrix $W$ the system reaches a balanced matrix in finite time with probability converging to $1$ as $n \rightarrow \infty$. They also characterized the converged value and its relationship to the initial value. 

In a recent paper~\cite{cisnerosvelarde2019structural}, Cisneros-Velarde et al. considered a \emph{pure-influence} model, where the self-appraisal (such as $w_{ii}$) is taken out of Equation~(\ref{eq:dynamic-w}) to be a more faithful interpretation of Heider's structural balance. 
They proved that when $W$ is symmetric their continuous-time dynamic model is exactly the gradient flow of an energy function called \emph{dissonance}~\cite{Marvel2009-lx}, defined as 
\begin{equation*}
    -\sum_{[ij],[jk],[ki] \in E} w_{ij} \cdot w_{jk} \cdot w_{ki}.
\end{equation*}
Dissonance characterizes the degree of violation to Heider's structural balance axioms in the current network. The global minimum of this energy function corresponds to signed networks that satisfy structural balance in the case of real-values appraisals. 
When the initial matrix $W$ is symmetric the authors also provided characterizations of the critical points of the dissonance function (aka the equilibrium states of the dynamic model).

\medskip
The discussions of opinion dynamics and dynamics with structural balance, so far, have focused on  node opinion changes or link appraisal changes, separately.  There is little work on combining both dynamics into a co-evolving model, which is the focus of this paper.

\section{Co-Evolution Model}

Suppose there are $n$ individuals, each one with its own opinion $v_i \in \mathbb{R}$. Define the opinion vector $V = (v_1, v_2, \dots, v_n )^{T} \in \mathbb{R}^{n}$. 
The influence model among the $n$ individuals is characterized as an $n\times n$ matrix $W = [w_{ij}]$ with entries taking real values. A positive value of $w_{ij}$ indicates a positive social influence between $i, j$, where the opinions under the influence become similar. A negative value of $w_{ij}$ means a negative influence and their opinions under influence become dissimilar. In our theoretical study, we consider the case of a complete graph.  The evolution model that we introduce works for any network. In our simulations we also evaluate networks and opinion co-evolution on a general graph. 

Both the opinions of individuals and the influence matrix are dynamically evolving. Assume that the initial opinion vector is $V(0)$ and the initial influence matrix is $W(0)$. In this paper we assume the initial weight matrix $W(0)$ is symmetric, i.e., $w_{ij}(0)=w_{ji}(0)$, $\forall i, j$.
Define the opinion vector and influence matrix at time $t$, in a discrete-time model, as $V(t)$ and $W(t)$ respectively. We propose the dynamic system governing the evolution of the relationship over integer time:
\begin{equation}\label{eq:discrete-model}
    \left\{
    \begin{array}{l}
        V(t+1) = V(t) + W(t)V(t) \\
        W(t+1) = W(t) + V(t) V(t)^T.
    \end{array}
    \right.
\end{equation}
In the first equation, the opinion of an individual $i$ is shifted by the weighted sum of its neighbors' opinions, with coefficients in the influence matrix $W$. In the second equation, the appraisal value $w_{ij}$ between two individuals $i, j$ is updated by the differences of opinions $v_i, v_j$. If $v_i, v_j$ generally agree (with a positive dot product), $w_{ij}$ moves in the positive direction; otherwise moves in the negative direction. 

In a continuous-time model, the dynamics are driven by the following ODE:
\begin{equation}\label{eqn:continuous-model}
    \left\{
    \begin{array}{l}
        V' = WV \\
        W' = VV^T.
    \end{array}
    \right.
\end{equation}
where $V'$ and $W'$ is the coordinate-wise time derivative of $V$ and $W$.

From this point on, we focus on solving the continuous time model. 
First we present a couple of basic properties of Equation~(\ref{eqn:continuous-model}). This means we can focus on solving the system defined by Equation~(\ref{eqn:continuous-model}) without losing generality. The detailed proof is provided in Appendix~\ref{subsec:model}.
\begin{restatable}{lemma}{conjugation}
\label{conjugation}
$ $
\begin{enumerate}
\item If $[V(t), W(t)]$ solves Equation~(\ref{eqn:continuous-model}) and $U$ is an orthogonal matrix, then $[U^TV, U^TWU]$ solves the same Equation~(\ref{eqn:continuous-model}) with initial condition $U^TV(0)$ and $U^TW(0)U$. In particular, $W(0)$ is symmetric if and only if $U^TW(0)U$ is.

\item If $W(0)$ is symmetric, then $W(t)$ remains symmetric for all $t$.

\item If $a,b>0$ are positive constants, then the equation
\begin{equation*} \label{modii}
 \left\{
    \begin{array}{l}
        V_1' = aW_1V_1 \\
        W_1' = bV_1V^T_1 
    \end{array}
    \right.
\end{equation*}
can be reduced to Equation~(\ref{eqn:continuous-model}) by taking $V=\sqrt{ab}V_1$ and $W=aW_1$.  
\end{enumerate}
\end{restatable}

The main objective of this paper is to analyze how this system evolves. In particular, we care about system evolution to reach \emph{sign stability} for $w_{ij}$, $\forall i, j$, and $v_i$, $\forall i$, as well as \emph{structural balance} --
$$\lim_{ t \to T} w_{ij}w_{jk}w_{ki} \geq 0,$$ for all indices $i,j,k$ where $[0, T)$ is the maximum interval on which the solution $W(t)$ exists.
Notice that in the classical structural balance theory, the two types of stable triangles -- with edge signs as either all positive ($+1$)
or have two negative ($-1$) and one positive -- satisfy this property.

The evolution of $W(t)$ and $V(t)$ is described in the following two lemmas.

\begin{lemma} \label{lemma22}
With the co-evolution model as in Equation~(\ref{eqn:continuous-model}), the dynamics of matrix $W$ follows the following Matrix Riccati Type Equation 
\begin{equation*}\label{eqn:Riccati1}
    W' = WW^T + C,
\end{equation*} 
where $C= V(0)V(0)^T - W(0)W(0)^T$ is a symmetric constant $n \times n$ matrix. If $W(0)$ is symmetric, then  $W(t)$  satisfies the Riccati equation
\begin{equation}\label{eqn:Riccati}
    W' = W^2 + C.
\end{equation}

\end{lemma}

\begin{proof}
We look at $W''$:
\begin{equation*}
    \begin{split}
         W'' & =  (W')' = (VV^T)' = V'V^T + V(V')^T  = WVV^T + V(WV)^T  = WVV^T + VV^TW^T  \\
      & =WW' + W'W^T = W(W^T)' + W'W^T = (WW^T)'.
    \end{split}
\end{equation*}
In the second last step, we use the equation $W'=(W^T)'$. This is because $W'(t)=V(t)V(t)^T$ is always symmetric. Thus, $W' = WW^T + C$, where $C$ is a constant matrix $C=W'(0)-W(0)W(0)^T =V(0)V(0)^T-W(0)W(0)^T$. Notice that $C$ is always symmetric. 

If $W(0)$ is symmetric, then $W(t)$ is always symmetric (by Lemma~\ref{conjugation} (2)) and $WW^T=W^2$.
\end{proof}

Remark that matrix $C$ in our setting is a special symmetric matrix. Specifically, $C+W(0)W(0)^T=V(0)V(0)^T$ has rank one. This property turns out to be useful for characterization of the system behavior. 

\begin{lemma}
The evolution of $V(t)$ satisfies
\begin{equation}\label{eqn:opinion_evolution}
    V''(t) = 2|V|^2\cdot V - C\cdot V
\end{equation}
\end{lemma}
\begin{proof}
Here, we use $W^2 = W' - C$ (Equation (\ref{eqn:Riccati})), $W' = VV^T$ and $ V' = WV$.
\begin{equation*} 
\begin{split}
V'' & = (V')'= (WV)'=  W'V+WV'= VV^TV+WWV  = |V|^2V+ W^2V = |V|^2V+ (W'-C)V \\  & = |V|^2V + (VV^T-C)V = 2|V|^2V-CV.
\end{split}
\end{equation*}
\end{proof}

\section{Analysis of the Opinion and Social Tie Evolution}

Our analysis has two parts. First we focus on the opinion evolution model (Equation (\ref{eqn:opinion_evolution})). Here we provide analysis of the asymptotic behavior for $V(t)$. Then we study the social tie evolution (Equation (\ref{eqn:Riccati})) for $W(t)$.  By solving Riccati equation explicitly for $W(t)$ we are able to provide more detailed characterization of the evolving behavior. 

\subsection{Analysis of Opinion Evolution}

By analyzing the opinion evolution (Equation (\ref{eqn:opinion_evolution})), our main result is the following.

\begin{theorem}\label{thm:main_theorem}
Let $[0, T)$ be the maximum interval of existence for the solution $(V(t), W(t))$ of the differential equation in Equation~(\ref{eqn:continuous-model}). For generic initial values $V(0)$ and $W(0)$,  either
\begin{enumerate}
    \item structural balance condition $\lim_{t \rightarrow T} w_{ij}w_{jk}w_{ki}>0, \forall i,j,k$ holds,  or 
    \item $T = +\infty$, $\lim_{t\rightarrow \infty}|V(t)| = 0$ and
    $\lim_{t\rightarrow \infty}V'(t) = 0$.
\end{enumerate}
Furthermore, in the first case, the normalized opinion vector $\frac{V(t)}{|V(t)|}$ converges,  i.e.,  $\lim_{t \to T} \frac{V(t)}{|V(t)|}$ exists. 
\end{theorem}

The theorem says that structural balance is always achieved, unless the opinions converge to a zero vector, in which case the entire network becomes neutral. We can consider the second case as a boundary case of structural balance. 

The rest of the subsection will focus on proving this theorem. An important observation is that the norm of $V$, $|V(t)|^2$, is a convex function. The detailed proof is in Appendix~\ref{app:opinion}.

\begin{restatable}{lemma}{lemmalength}
\label{lemma:length}
The length function $\varphi(t):= V^T V = |V(t)|^2$ is strictly convex and $\varphi''(t)>0$ unless $V(t)\equiv 0$.
\end{restatable}
Now, let us understand Equation (\ref{eqn:opinion_evolution}) using coordinates. Since the matrix $C$ is symmetric, by the orthogonal diagonalization theorem, there exists an orthogonal matrix
$$U = [\beta_1, \cdots, \beta_n]=[u_{ij}]_{ n \times n}$$
such that
$C\beta_i = a_i \beta_i, i=1, 2, \cdots, n,$
where  $a_1, \cdots, a_n$ are eigenvalues of $C$. 
By our assumption that $C = V(0)V(0)^T - W(0)W(0)^T$, the eigenvalues $a_1, \cdots, a_n$ are non-positive except for one. So we may assume $a_1 \geq 0$ and $a_2, \cdots, a_n \leq 0$. Because $\beta_1, \cdots, \beta_n$ form an orthonormal 
basis of $\mathbb{R}^n$, we can write
\begin{equation*}
    V(t) = \sum_{i=1}^n \lambda_i(t)\beta_i = U\lambda, \mbox{ where } \lambda = [\lambda_1, \cdots, \lambda_n]^T.
\end{equation*}
This implies that $V'(t) = \sum_{i=1}^n \lambda_i'(t)\beta_i$, $V''(t) = \sum_{i=1}^n\lambda_i''(t) \beta_i$, $|V(t)|^2 = \sum_{j=1}^n \lambda_j^2(t)$ and $C\cdot V = \sum_{i=1}^n a_i \lambda_i \beta_i$.

Therefore, Equation (\ref{eqn:opinion_evolution}) becomes
\begin{equation*}
    \sum_{i=1}^{n} \lambda_i''(t) \cdot \beta_i = \sum_{i=1}^n (2(\sum_{k=1}^n \lambda_k^2) - a_i) \lambda_i \cdot \beta_i
\end{equation*}
Since $\beta_1, \cdots, \beta_n$ are independent, we obtain the system of ODE with $\lambda = [\lambda_1, \cdots, \lambda_n]^T$ in the form
\begin{equation}\label{eqn:lambda}
    \lambda_i''(t) = (2\sum_{k=1}^n \lambda_k^2 - a_i) \lambda_i(t), i=1, 2, \cdots, n.
\end{equation}

Denote $W = [w_{ij}(t)]$. Then $W' = V\cdot V^T$ implies
$$W' = U\cdot \lambda \cdot \lambda^T \cdot U^T$$
Therefore, $w'_{ij}(t)=\sum_{k,l=1}^n u_{ik}u_{jl}\lambda_{k}\lambda_l$ and $w_{ij}(t) = \sum_{k,l=1}^n (\int_0^t \lambda_k(s)\cdot \lambda_l(s) ds) u_{ik}u_{jl} + w_{ij}(0)$.

Our next goal is to show the following proposition,

\begin{restatable}{proposition}{propgrowth}
\label{prop:growth}
If $T < + \infty$ or if $T = + \infty$ and $\lim_{t\rightarrow T} |V(t)| = L > 0$, then there exists one term $\psi_{hh}$ among $\psi_{kl}(t) = \int_{0}^t \lambda_k(s) \lambda_l(s) ds$ which has the maximum growth rate as $t \rightarrow T$ and $\lim_{t\rightarrow T} \psi_{hh}(t) = + \infty$.
\end{restatable}

Assuming the proposition~\ref{prop:growth}, Theorem~\ref{thm:main_theorem} follows. Indeed, the leading term in $w_{ij}(t)$ is $u_{ih}u_{jh}\psi_{hh}$ as $t \rightarrow T$. Therefore the sign of $w_{ij}$ is the same as the sign of $u_{ih}u_{jh}\psi_{hh}$. The leading term of $w_{ij}w_{jk}w_{ki}$ as $t \to T$ is 
$$ u_{ih}^2 u_{jh}^2 u_{kh}^2 \psi_{hh}^2 \geq 0.$$ 
This shows that structural balance occurs eventually for generic initial values. Here the generic condition is used to ensure that all entries $u_{ij}$ of the orthogonal matrix $[u_{ij}]$ are not zero and $\psi_{hh}$ is the unique term with the maximum growth rate. Finally, if $T=\infty$ and $\lim_{t \to \infty} |V(t)|=0$, then by Corollary \ref{cor:v'}, $\lim_{t \to \infty} V'(t)=0$.

The proof for proposition~\ref{prop:growth} is fairly technical and can be found in Appendix~\ref{app:opinion}.

\subsection{Analysis of Social Tie Evolution}

The analysis on the evolution of $V$ in the previous section shows convergence. To further understand the community formed at the limit of convergence, we need to study the evolution of $W$. 
The following theorem explains the reason behind the appearance of structure balance when at least one eigenvalue of $W(t)$ tends to infinity. This was proved in~\cite{Marvel2009-lx}. We include the statement here and the proof in Appendix~\ref{subsec:structural} for completeness. 
\begin{restatable}[\cite{Marvel2009-lx}]{theorem}{thmconvergence}
\label{thm:convergence}

\label{kl} Suppose $W(t)$, $t \in [0, T)$,  is a continuous family of symmetric matrices 
such that
\begin{enumerate}
    \item $W(t)$ has a unique largest eigenvalue, denoted by $\beta_1(t)$, which tends to infinity as $t \to T$, 
    \item  all eigenvectors of $W(t)$ are time independent, and
    \item  all components of the $\beta_1(t)$ eigenvector are not zero.
\end{enumerate}
Then
$$ w_{ij}w_{jk}w_{ki} > 0,\, \forall i, j, k, i\neq j, j\neq k, i\neq k,$$
for all time $t$ close to $T$, i.e.,  the structural balance of the whole graph is satisfied. 

If (2) does not hold, we have $$ w_{ij}w_{jk}w_{ki} \geq 0,  \forall i, j, k, i\neq j, j\neq k, i\neq k. $$
\end{restatable}

Furthermore, the two antagonistic communities are given by $U^+=\{i \in V|u_i>0\}$ and $U^-=\{i \in V| u_i<0\}$ where $u=[u_1, ..., u_n]^T$ is a $\beta_1(t)$ eigenvector. All edges connecting vertices within the same community are positive while edges connecting two vertices in different communities are negative. When one of $U^+, U^-$ is empty, the  network has only one community.

The model in  \cite{Marvel2009-lx} is the Riccati equation $W'=W^2$, whose solution is $W(t)=W(0)(1-W(0)t)^{-1}$. As a consequence, Marvel et. al.  \cite{Marvel2009-lx} showed that structure balance occurs in the Riccati equation $W'=W^2$ for generic initial parameter $W(0)$ with a positive eigenvalue at finite time. Our model strictly generalizes the previous model and consider how the initial opinions may influence the system evolution. In the following we show how to solve the general Riccati equation and also when an eigenvalue goes to infinity. 

With most of the details in the Appendix, we carry out rigorous analysis of the general form of the matrix Riccati equation as stated below.
\begin{equation}
\label{eqn:Evolution}
    \left\{
    \begin{array}{rl}
        W' & = W^2 + C \\
        W(0) & = B.
    \end{array}
    \right.
\end{equation}
For general matrices $B, C$ we can solve for $W(t)$ as shown in the following theorem. The proof details can be found in Appendix~\ref{subsec:general-Riccati}. 

\begin{restatable}{theorem}{mainth}
\label{mainth}
The solution $W(t)$ is given by the explicit formula that $W(t) = -Z(t)\cdot Y(t)^{-1}$, where
\begin{eqnarray}
Y(t) &= & \sum_{n=0}^{\infty} \frac{(-1)^n t^{2n} C^n}{(2n)!} + \sum_{n=0}^{\infty}\frac{(-1)^{n+1}t^{2n+1}C^nB}{(2n+1)!} \\
Z(t) &=&  \sum_{n=0}^{\infty}\frac{(-1)^{n+1}t^{2n}C^nB}{(2n)!} +  \sum_{n=0}^{\infty} \frac{(-1)^{n+1} t^{2n+1} C^{n+1}}{(2n+1)!}.
\end{eqnarray}
\end{restatable}

In our co-evolution model, $B=W(0)$ is assumed to be symmetric. Thus $C$ is also symmetric. This allows us to simplify the solution further, as shown in Appendix~\ref{subsec:socialtie}.
Then we analyze a special case when $BC=CB$.  In this case we can characterize the conditions when structural balance is guaranteed to occur, with details in Appendix~\ref{subsec:structural}, \ref{subsec:BCequalCB}, and \ref{subsec:structural-commute}.
Specifically, using a basic fact that two commuting symmetric matrices can be simultaneously orthogonally diagonalized~\cite{Hoffman1971-bc}, we can get the following theorem where the conditions of the eigenvalues of $B, C$ for structural balance are characterized:

\begin{restatable}{theorem}{thmmain}
\label{thm:main}
Suppose $W(t)$ solves the Riccati equation $W'=W^2+C$, $W(0)=B$ where $B, C$ are symmetric with $BC=CB$.
Then eigenvalues of $W(t)$ converge to elements in $(-\infty, \infty]$ as $t \to T;$ meanwhile there is sign stability, i.e., $\lim_{t \to T} w_{ij}(t) \in [-\infty, \infty]$ exists for all $i,j$. 
If $U$ is an orthogonal matrix such that 
\begin{equation*}
    \begin{split}
        U^TCU & = \diag(a_1^2, \cdots, a_k^2, -d_1^2, \cdots, -d_l^2, 0, \cdots, 0)\\
        U^TBU & = \diag(\lambda_1, \cdots, \lambda_k, \mu_1, \cdots, \mu_l, \delta_1, \cdots, \delta_h).
    \end{split}
\end{equation*}
where $a_i, d_j > 0$, 
then $W(t)$ is given by the following explicit function, \begin{equation}\label{eqn:W}
\begin{aligned}
      W(t)  = & U\cdot \diag(\frac{a_1 \sin{(a_1 t)} + \cos{(a_1 t)}\lambda_1}{\cos{(a_1 t)}-\frac{1}{a_1}\sin{(a_1 t)}\lambda_1}, \cdots, \frac{a_k \sin{(a_k t)}) + \cos{(a_k t)}\lambda_k}{\cos{(a_k t)}-\frac{1}{a_k}\sin{(a_k t)}\lambda_k},\\
       & -\frac{d_1 \sinh{(d_1 t)}-\cosh{(d_1 t)}\mu_1}{\cosh{(d_1 t)} - \frac{1}{d_1}\sinh{(d_1 t)}\mu_1}, \cdots, -\frac{d_l \sinh{(d_l t)}-\cosh{(d_l t)}\mu_l}{\cosh{(d_l t)} - \frac{1}{d_l}\sinh{(d_l t)}\mu_l}, \\
       & \frac{\delta_1}{1 - t\delta_1}, \cdots, \frac{\delta_h}{1 - t\delta_h}) \cdot U^T.
\end{aligned}
\end{equation}
Further, structural balance 
 $$ \lim_{t \to T} w_{ij}w_{jk}w_{ki} \geq 0$$
occurs in the finite time for $W(t)$ if  $W(t)$ has an unique largest eigenvalue and one of the following conditions holds:
\begin{enumerate}
    \item There exists some $\delta_i>0$,
    \item There exists some $ a_i>0$,
    \item there exists some $\mu_i > d_i$.
\end{enumerate}
\end{restatable}

In our co-evolution model, $BC=CB$ happens when $V(0)$ is a eigenvector of $W(0)$, an interesting initial condition. 
To see that, recall $B=W(0)$, $C= V(0)V(0)^T - W(0)W(0)^T$, and $W(0)=W(0)^T$. To check if $BC=CB$, we just need to check if $W(0)V(0)V(0)^T=V(0)V(0)^TW(0)$ and apply the following Lemma (Proof in Section~\ref{subsec:structural-commute}). 

\begin{restatable}{lemma}{lemmaBCcommute}
\label{lemma:BCcommute}
Suppose $A$ is a symmetric matrix and $v$ is a non-zero column vector. Then 
$A vv^T=vv^TA$ is equivalent to $Av=\alpha v$, i.e., $v$ is an eigenvector of $A$.
\end{restatable}

Further, the equation in our co-evolution model, i.e.,  Equation~(\ref{eqn:Riccati}), satisfies $C+B^2=V(0)V(0)^T$. Notice that the right-hand side $V(0)V(0)^T$ is an $n
\times n$ matrix with rank one. This property actually ensures that the conditions characterized in Theorem~\ref{thm:main} are met and thus structural balance is guaranteed. At the same time, the convergence rate is $O(\frac{1}{|T - t|})$, which is proved by Lemma~\ref{lemma:sin} in Appendix~\ref{subsec:BCequalCB}.

\begin{restatable}{corollary}{corcomute}
\label{cor:comute}
For Equation~(\ref{eqn:continuous-model}), if $V(0)\neq 0$ is an eigenvector of $W(0)$, then $V(t)$ remains to be an eigenvector of $W(t)$ for all $t$ and structural balance must occur in finite time for $W(t)$.
\end{restatable}

If $V(0)=0$, the system stays at the fixed point with $V$ remaining zero and the weight matrix unchanged. 

The case when $V(0)$ is an eigenvector of $W(0)$ includes a few interesting cases in practice. When $W(0)=0$ or $W(0)=I$, this models a group of individuals that start as complete strangers with uniform self-appraisals. Their non-homogeneous initial opinions $V(0)$ may drive the network to be segmented over time. Finally the fact that $V(t)$ remains an eigenvector of $W(t)$ for all $t$ follows from Equation (\ref{eqn:WWW}) in Appendix C.

We conjecture that even in the general case (when $V(0)$ is not necessarily an eigenvector of $W(0)$) the limit vector $\lim_{t \to T} V(t)/|V(t)|$ is an eigenvector of the limit tie relation matrix $\lim_{t \to T} W(t)/|V(t)|$.  This is supported by our numerical evidences and Corollary \ref{cor:wv} which says that $\lim_{t \to T} V(t)/|V(t)|$ is an eigenvector of $\lim_{t \to T} (W(t)/|V(t)|)^2$.

\section{Simulations}
In this section, we provide simulation results to accompany our theoretical analysis of the co-evolution model. 
We also present simulation results on general graphs and real world data sets to understand the behavior of network evolution.

\begin{figure}[h!]
    \centering
    \includegraphics[width=0.9\columnwidth]{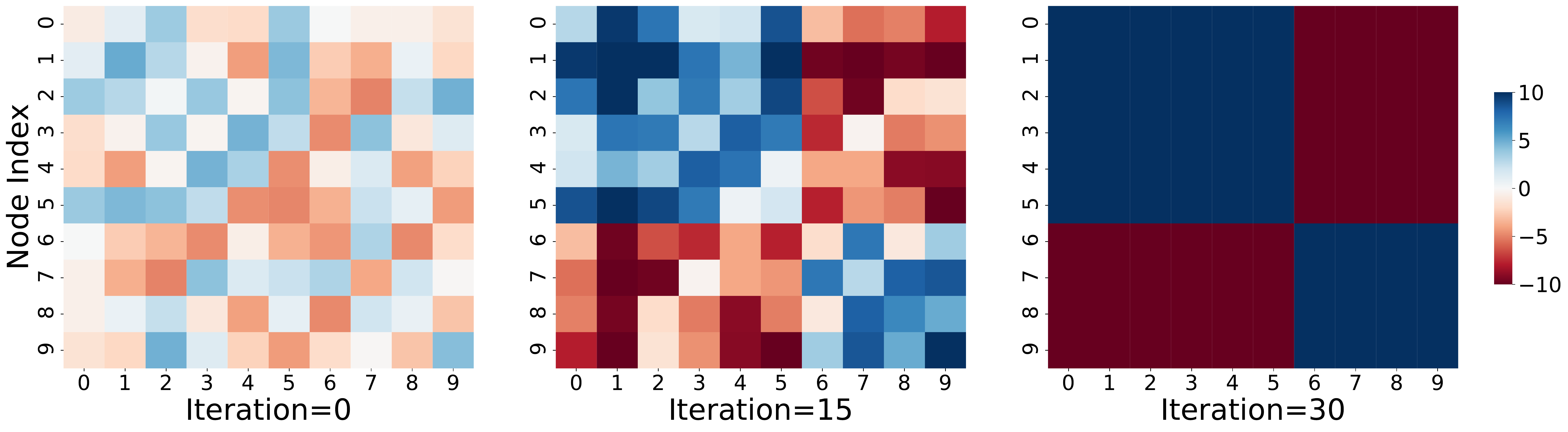}
    \caption{The weight matrix in evolution. The $x$ and $y$ axis show vertex indices. The cell at $(i, j)$ represents the edge weight between node $i$ and node $j$ with color showing edge weight. The first plot is the initial edge weights, which are assigned random values. The second plot shows the weight matrix after $15$ iterations. Patterns start to show up with two diagonal blocks showing positive values and the off-diagonal blocks with negative values. The last plots is the weight matrix at convergence where the first $6$ nodes are in one community and the other nodes are in a different community. Edges within communities have positive weights and edges between communities have negative weights. }
    \label{fig:heatmap}
\end{figure}

\begin{figure}[h!]
\centering
\subfigure[Initial Karate club graph]{
\centering
\includegraphics[width=0.35\columnwidth]{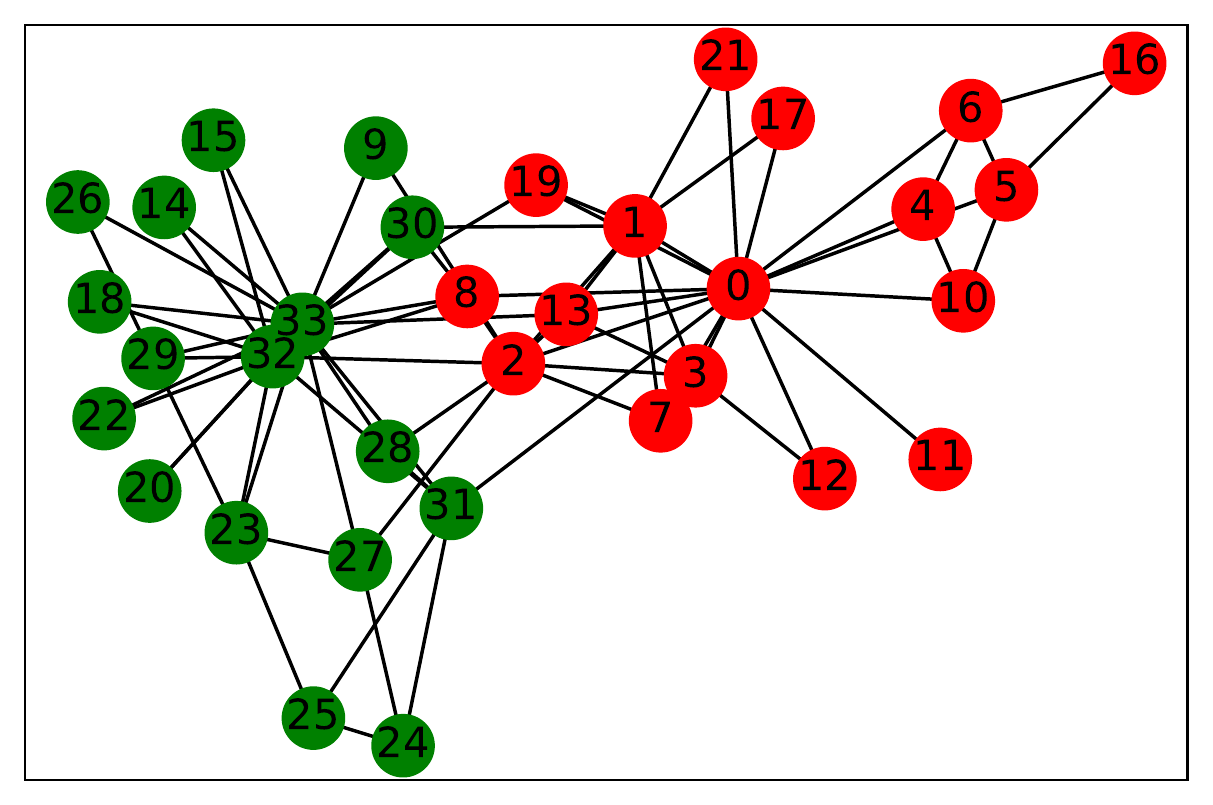}
\label{fig:karate_1}
}
\subfigure[Separated communities results]{
\centering
\includegraphics[width=0.35\columnwidth]{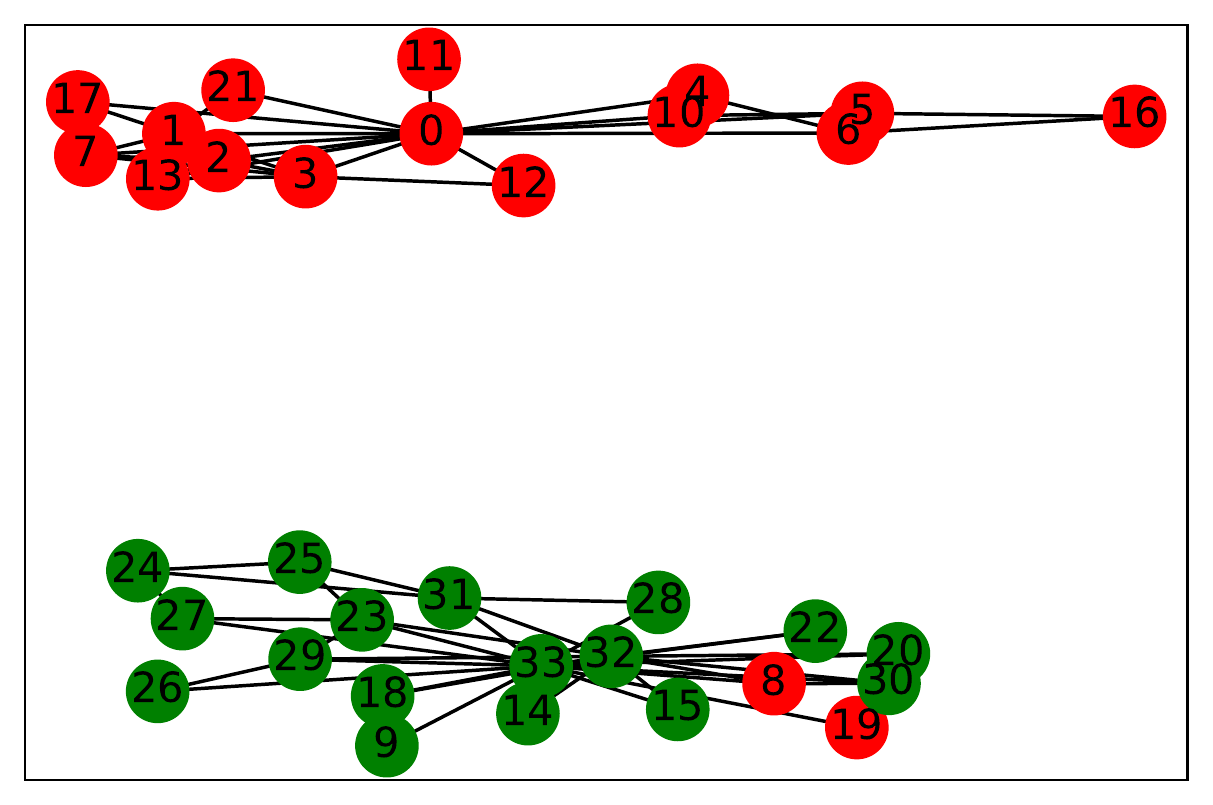}
\label{fig:karate_2}
}
\subfigure[Initial political blogs network]{
\centering
\includegraphics[width=0.28\columnwidth]{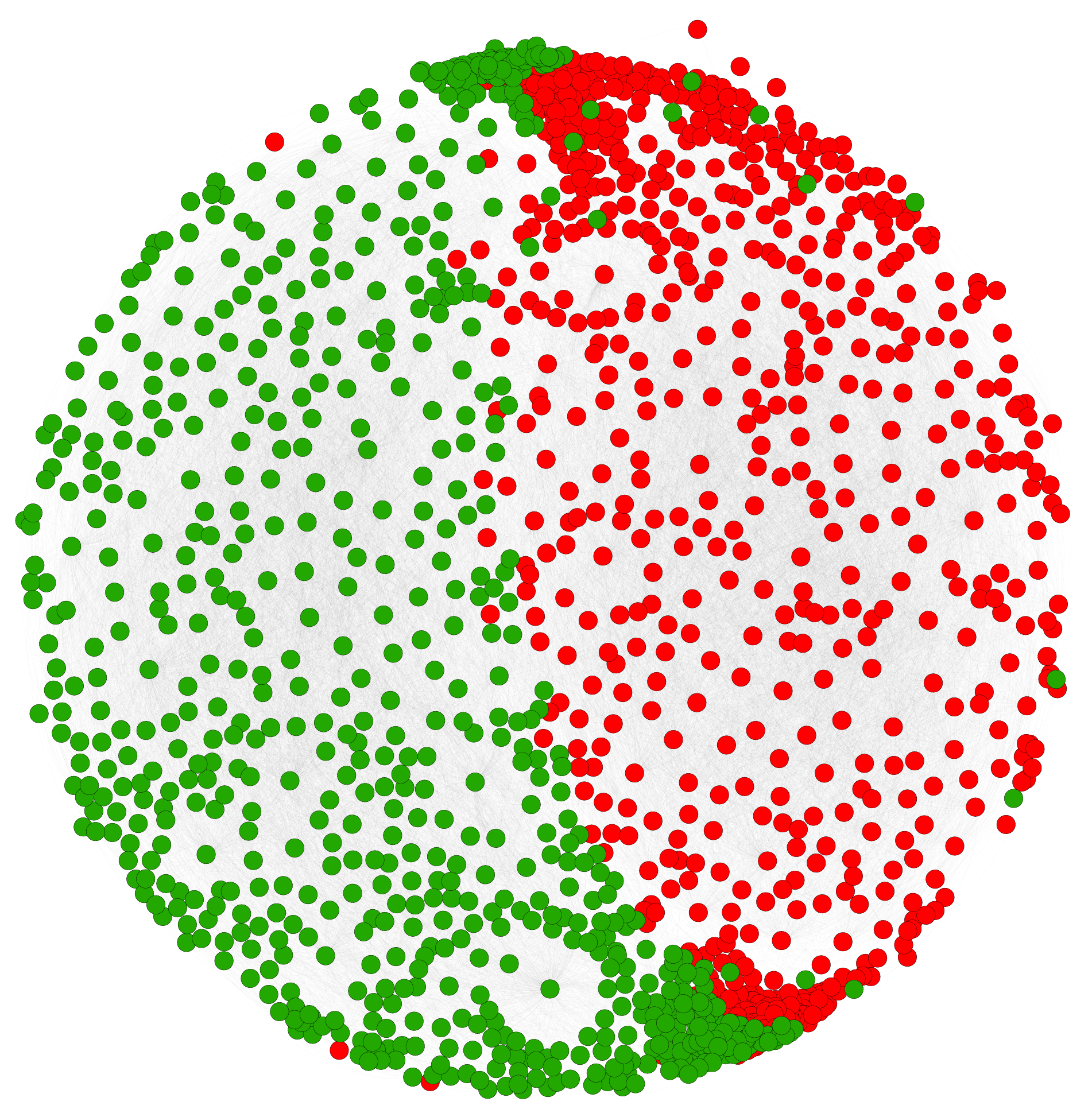}
\label{fig:blog_1}
}
\hspace{1cm}
\subfigure[Separated communities results]{
\centering
\includegraphics[width=0.36\columnwidth]{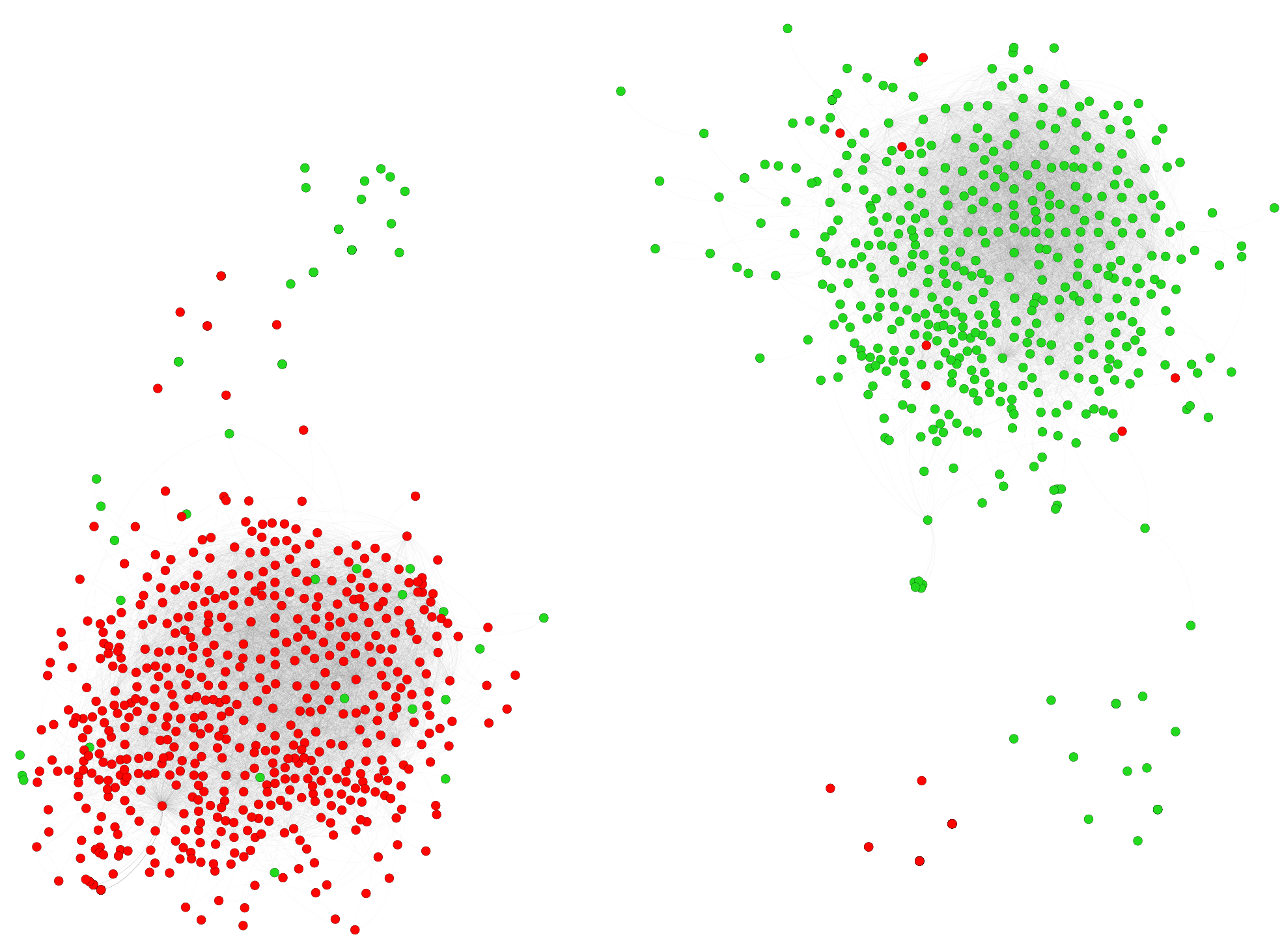}
\label{fig:blog_2}
}
\caption{Community detection in two real-world data sets. The color of vertices represents their ground truth opinions. Initially, only a small number of seeds are assigned their true opinions and other nodes start with initial opinions of $0$. All edges start with a small positive weight. As the network converges, the edges with negative weights are removed. The communities discovered as shown in the right column faithfully reflect the ground truth. In the first experiment (a, b), the node $\#0$ and the node $\#33$ are assigned opposite initial opinions. It nearly predicts the same division as in the ground truth except for two nodes $\#8$ and $\#19$ which are somewhat ambiguous. In the second experiment (c, d), $20\%$ random nodes are assigned initial opinions according to the ground truth. The prediction achieves an average accuracy of $97.21\%$ compared with the ground truth.}
\label{fig:application}
\end{figure}

Here is a brief summary of observations from simulations. The details can be found in the Appendix. 
\begin{itemize}
\item Except a few carefully crafted cases\footnote{For example, if $W(0) = B =  \diag(-\frac{1}{a}, -\frac{1}{a}, \cdots, -\frac{1}{a})$ and we use the dynamic update rule $v_i(1) = v_i(0) + a\cdot w_{ii}(0)v_i(0) = 0$, for all $1 \leq i \leq n$. After one iteration, all node opinions become $0$. After that, the opinions and weights do not change anymore. } that make the opinion vector to be zero, strict structural balance is always reached, regardless of whether the graph is complete or general, whether $B, C$ commute or not, or whether $W(0)$ is symmetric or as a general matrix. An example is shown in Figure~\ref{fig:heatmap}.
\item In general, we observed through simulations that the number of iterations to convergence is inversely proportional to the magnitude of initial opinions and the edge density in the graph. 
\item In two real world data sets (the Karate club and a political blog data set), a few initially planted polarized opinions can successfully predict the ground truth community structure with high accuracy. See Figure~\ref{fig:application}.
\end{itemize}


\section{Conclusion and Future Work}
In this paper, we have provided a co-evolution model for both opinion dynamics and appraisal dynamics. We provided solutions to the system and rigorously characterized how the stable states depend on the initial parameters.


There are a few follow-up problems that remain open, for example, 
when the social ties are directional/asymmetric, when the network is not a complete graph, and when each agent has an $m$-dimensional opinion vector. We include some discussion and conjectures on these cases in Appendix~\ref{subsec:general} and consider this as interesting future work. 

\newpage
\bibliographystyle{ACM-Reference-Format}
\bibliography{evolution,community}

\appendix
\input{model}
\input{newmath}

\input{Riccati}
\input{extension}

\input{simulation}
\end{document}

%% file: model.tex
\section{Model}~\label{subsec:model}

\conjugation*
\begin{proof}

Part (1) follows from standard computation and that $UU^T=U^TU=I$ for an orthogonal matrix $U$. Also, a matrix $A$ is symmetric if and only if $U^TAU$ is symmetric.

For part (2), note that the equation $W'=VV^T$ implies that $W'(t)$ is always symmetric, i.e., $W'(t) = W'(t)^T$. Now if $W(0)$ is symmetric, then $W(t)$ and $W^T(t)$ are solutions of the same differential equation $W'(t) =W'(t)^T$ with the same initial value. By the uniqueness theorem of the solution of ordinary differential equation, $W(t)=W^T(t)$.

Part (3) follows by rewriting Equation~(\ref{modii}) as
\begin{equation*}
 \left\{
    \begin{array}{l}
    (\sqrt{ab}    V_1)' = (aW_1) (\sqrt{ab}    V_1)  \\
       (aW_1)' = (\sqrt{ab}    V_1)(\sqrt{ab}V_1)^T.
    \end{array}
    \right.
\end{equation*}
Take $V=\sqrt{ab}V_1$ and $W=aW_1$. This becomes Equation~(\ref{eqn:continuous-model}).
\end{proof}

%% file: newmath.tex
\section{Evolution of Opinion Dynamics}\label{app:opinion}
\lemmalength*
\begin{proof}
Recall that $\varphi(t)=|V(t)|^2$. 

$$\varphi'(t) = (V')^TV + V^TV' = (WV)^TV+V^TWV = 2V^TWV.$$
Therefore,
\begin{equation}\label{eqn:length}
    \begin{split}
        \varphi''(t) & = 2((V')^TWV + V^TW'V+ V^TWV') = 2((WV)^TWV + V^TVV^TV + V^T WWV)\\
        & = 2(|WV|^2 + |V|^4 + V^TW^TWV) = 2(2|WV|^2 + |V|^4)  = 2(2|V'(t)|^2 + |V(t)|^4 ) \geq 0
    \end{split}
\end{equation}
Now if $\varphi''(t_0) = 0$, then $V'(t_0) = V(t_0)=0$ by Equation (\ref{eqn:length}).
This shows that $V(t)$ is the solution of the ODE (Equation (\ref{eqn:opinion_evolution})) with the initial condition $V(t_0) = V'(t_0) = 0$. But $0$ is also the solution. Therefore, by the uniqueness of solution of ODE with initial value, $V(t)\equiv 0$. This ends the proof.
\end{proof}

\begin{corollary} \label{cor:v'} If $V(t)$ solves Equation (\ref{eqn:opinion_evolution}), then  
\begin{equation} \label{v'}
|V'(t)|^2 =|V(t)|^4 -V^TCV. \end{equation}
In particular, if $\lim_{t \to T} V(t)=0$, then $\lim_{t \to T} V'(t)=0.$
\end{corollary}

Indeed, by $(|V(t)|^2)'=2V^TV'$ and Equation (\ref{eqn:opinion_evolution}) that $V''=2|V|^2V-CV$, we have $(|V(t)|^2)''=2 ( V')^T V' + 2V^T V'' = 2|V'(t)|^2 + 2V^T(2|V|^2V-CV) = 2|V'(t)|^2 +4|V|^4-2V^TCV.$  Comparing it with  Equation  (\ref{eqn:length}), we see the corollary holds. The last statement of the corollary follows from Equation (\ref{v'}).

We now prove Proposition~\ref{prop:growth} using several lemmas. For simplicity, if $f(t)$ is a function defined on an open internal $(a, b)$ (here $b$ may be $+\infty$), we say $f$ has property $P$ (e.g., positive, non-negative, monotonic, convex etc) near $b$ if there exists $\epsilon > 0$ such that the restriction of $f$ on the interval $(b-\epsilon, b)$ (if $b < +\infty$) or $(\frac{1}{\epsilon}, \infty)$ (if $b=+\infty$) has property $P$. For example, $t^4 + 3t^3 - 9 t^2 +t - 5$ is positive and convex near $+\infty$. The notation $C^{k}(a,b)$ stands for all functions $f(t)$ for which $f, f', \cdots, f^{(k)}$ are continuous on the interval $(a, b)$. 

\begin{lemma}\label{lemma:f_property}
Suppose $f \in C^2(a, b)$ and $p(t)>0$ on $(a, b)$ such that 
\begin{equation}\label{eqn:f''}
    f''(t) = p(t)f(t) \mbox{ on }(a, b)
\end{equation}
and $f(t)$ is not identically zero on any sub-interval. Then
\begin{enumerate}
    \item $f$ has at most one root in $(a, b)$.
    \item $f$ has the same sign near $b$ (i.e., always positive or negative).
    \item $f$ is monotonic near $b$.
    \item the limit $\lim_{t \to b^-} f(t)$ exists (the limit may be $\pm \infty)$).
    \end{enumerate}
Furthermore, if $g''(t) \geq q(t)g(t)$ and $q\geq 0$ on $(a, b)$, then either $g(t) > 0$ near $b$ or $g(t) \leq 0$ near b.
\end{lemma}
\begin{proof}
To see part (1), if $f$ has two roots in $(a,b)$, then since $f$ is not identically zero on any interval, there exist two adjacent roots $f(c) = f(d)$ where $c < d$ and $f$ has no roots in the open interval $(c, d)$. By replacing $f$ by $-f$ if necessary, we may assume that the restriction function $f|_{(c, d)} > 0$. Then by Equation~(\ref{eqn:f''}), $f''(t) = p(t)f(t)>0$ on $(c, d)$. Therefore $f|_{[c,d]}$ is a convex function which has two minimum values at $c, d$. This implies that $f|_{[c, d]}\equiv 0$ which contradicts the assumption.

Part (2) follows from the part (1) easily.

To see part (3), we first show that $f'(t)$ has at most two roots in the open interval $(a, b)$. Suppose otherwise that $f'(t)$ has three roots in $(a, b)$. Then by the Mean Value Theorem, $f''(t)$ has two roots in $(a, b)$, one in each interval bounded by roots of $f'(t)$. 
But $f''(t)=p(t)f(t)$ with $p(t)>0$ says $f$ and $f''$ have the same roots. This implies that $f$ has two roots in $(a, b)$ which contradicts the part (1). Since $f'$ has only two roots, it follows that $f'(t) > 0$ near $b$ or $f'(t) < 0$ near $b$. Therefore $f$ is monotonic near $b$.

Part (4) follows from the well-known theorem that if $h(t)$ is monotonic in an open interval $(a, b)$, then the limit $\lim_{t\rightarrow b^{-}} h(t)$ always exists (limit value of the limit may be $\pm \infty$).

Finally, to prove the last statement, we consider two cases. In the first case, there are no sequence $\{r_m \}$ of roots of $g$ such that $\lim_{m \to \infty} r_m = b$. Then clearly $g(t) > 0$ or $g(t) <0$ near b. In the remaining case, we have an increasing sequence of roots, $r_1 < r_2< \cdots < r_m < \cdots$ of $f$ such that $\lim_{m\rightarrow \infty} r_m = b$. We claim that $g|_{[r_1, b)}\leq 0$. Suppose otherwise that $g(t_0) > 0$ for some $t_0 \in (r_1, b)$.
Let $c$ (respectively $d$) be the largest (respectively smallest) root of $g$ such that $c < t_0$ (respectively $d > t_0)$. By the assumption, both $c$ and $d$ exist. Furthermore $c < t_0 < d$ and $g$ has no root in the interval $(c, d)$. Therefore, due to $g(t_0) > 0$, $g|_{(c, d)} > 0$. By the condition $g'' = qg$, with $q \geq 0$, we see that $q''|_{(c, d)} \geq 0$. Therefore $g(t)$ is convex on $[c, d]$ and has two minimum values $0(=g(c) = g(d))$. But that implies $g|_{[c, d]} = 0$ and contradicts $g(t_0) > 0$. 
\end{proof}

\begin{corollary}\label{corollary:sol}
Suppose $\lambda(t) = [\lambda_1(t), \cdots, \lambda_n(t)]^T$ solves the ODE Equation (\ref{eqn:lambda}) and $a_1 \geq 0$, $a_2, \cdots, a_n \leq 0$ on the maximum interval $[0, T)$. Then 

\begin{enumerate}
    \item $\lim_{t\rightarrow T} |V(t)|^2 = \lim_{t\rightarrow T} \sum_{k=1}^{n} \lambda_k^2(t)$ exists.
    \item For all $i$,  $\lim_{t\rightarrow T} \lambda_i^2(t)$ exists.
    \item Assuming that $\lim_{t \to T} |V(t)|^2=\infty$, for $i\neq j$, either $\lambda_i^2(t) \geq \lambda_j^2(t)$ near $T$ or $\lambda_i^2(t) \leq \lambda_j^2 (t)$ near $T$.
    \item  For $\lim_{t \to T} |V(t)|^2 <\infty$, then all limits $\lim_{t \to T} \lambda_k^2(t)$ are finite and can be ordered.
\end{enumerate}
\end{corollary}

\begin{proof}
For (1), by Lemma~\ref{lemma:length}, $\phi(t) = |V(t)|^2$ is convex in $[0, T)$. Hence $\phi(t)$ is monotonic near $T$ and $\lim_{t\rightarrow T}\phi(t)$ exists.

For (2), let us assume $V(t)$ is not identically zero. Otherwise the result holds trivially. If $i \geq 2$, then by Equation~(\ref{eqn:lambda}), $a_i \leq 0$ and Lemma~\ref{lemma:length}, we have $2|V(t)|^2 - a_i > 0$ on $[0, T)$ and $\lambda_i''(t) = p_i(t)\lambda_i(t)$, where $p_i(t) = 2|V(t)|^2-a_i > 0$. Therefore by Lemma \ref{lemma:f_property}, $\lim_{t\rightarrow T}\lambda_i^2(t)$ exists. If $i = 1$, we have $\lambda_1^2(t) = |V(t)|^2 - \sum_{k=2}^n \lambda_k^2(t)$. 
Now if $\lim_{t\rightarrow T}|V(t)|^2 < +\infty$, then $\lim_{t \rightarrow T} \lambda_k^2(t) < + \infty$, for $k = 2, \cdots, n$. Therefore, $\lim_{t \rightarrow T} \lambda_1^2(t) =\lim_{ t \to T}(|V(t)|^2 -\sum_{k_2}^n \lambda_{k}^2)$
exists and is finite. If $\lim_{t \rightarrow T}|V(t)|^2 = + \infty$, then $2|V(t)|^2 - a_1 > 0$ for $t$ near $T$. The equation $\lambda_1''(t) = (2|V(t)|^2 - a_1)\lambda_1(t)$ is of the form $\lambda_1''(t) = p_1(t) \lambda_1(t)$ where $p_1(t) > 0$ near $T$. Therefore by Lemma~\ref{lemma:f_property}(4), $\lim_{t\rightarrow T}\lambda_1^2(t)$ exists.

For (3), since $\lambda_i^2 \geq \lambda_j^2$ is the same as $|\lambda_i|\geq |\lambda_j|$ and if $\lambda_i$ solve Equation~(\ref{eqn:lambda}) so is $-\lambda_i(t)$. The assumption that $\lim_{t \to T} |V(t)|^2 =\infty$ implies $|V(t)|^2-a_k \geq 0$ for $t$ near $T$. We may assume, using Lemma~\ref{lemma:f_property}(2), that $\lambda_i(t) \geq 0$ and $\lambda_j(t)\geq  0$ near $T$. Our goal is to show, under the assumption that $\lambda_i, \lambda_j \geq 0$ near $T$, either $\lambda_i \geq \lambda_j$ or $\lambda_i \leq \lambda_j$ near $T$. Without loss of generality, we may assume that $a_j \geq a_i$ and $a_i \leq 0$ (Note $a_1 \geq 0$ and $a_2, \cdots, a_n \leq 0$). Then using Equation~(\ref{eqn:lambda}), we have
\begin{equation*}
    \begin{split}
        (\lambda_i - \lambda_j)'' & = \lambda_i'' - \lambda_j'' = (2|V(t)|^2 - a_i)  \lambda_i - (2|V(t)|^2 - a_j)  \lambda_j \\
        & = (2|V(t)|^2 - a_i)(\lambda_i - \lambda_j) + (a_j - a_i)\lambda_j \geq (2|V(t)|^2 - a_i) (\lambda_i- \lambda_j).
    \end{split}
\end{equation*} 
Since $a_j - a_i \geq 0$ and $\lambda_j \geq 0$. Now due to $a_i \leq 0$, $2|V(t)|^2 - a_i \geq 0$. Therefore $(\lambda_i -\lambda_j)'' \geq q(t)(\lambda_i - \lambda_j)$ when $q \geq 0$ near $T$. By the last proof of Lemma~\ref{lemma:f_property}, we see either $\lambda_i > \lambda_j$ or $\lambda_i \leq \lambda_j$ near $T$. This ends the proof.

Part (4) follows from part (2) and the assumption which implies $\lim_{t \to T} \lambda_k^2$ are finite real numbers.  Therefore, we can order them. 
\end{proof}


Now, let us prove Proposition~\ref{prop:growth}. 
\propgrowth*

\begin{proof}

There are two cases depending on the maximum interval of existence $[0, T)$ being finite, i.e., $T < +\infty$ or infinite $[0, + \infty)$, i.e., $T = + \infty$.

\medskip
\textbf{Case 1.} $T < + \infty$. Recall the basic global existence of solution to ODE~\cite{Walter1998ODE}.

\begin{theorem}~\label{thm:differential}
(Existence) Suppose $F(t, x) \in C^{1}(\mathbb{R}\times \mathbb{R}^{m})$ and $[t_0, T)$ is the maximum interval of existence of the solution $x(t)$ to $x'(t) = F(t, x(t))$ with $x(t_0) = x_0$. Then the path $\{(t, x(t)) | t \in [t_0, T) \}$ does not lie in any bounded set in $\mathbb{R} \times \mathbb{R}^{n}$.
\end{theorem}

Now, for $T < +\infty$, Theorem~\ref{thm:differential} implies $\lim_{t\rightarrow T} |V(t)| = +\infty$. Indeed, if $\lim_{t\rightarrow T} |V(t)| < + \infty$, then $V(t)$ is bounded on $[0, T)$. This implies $W'(t) = V(t)\cdot V(t)^T$ is bounded. but $W(t) = W(0) + \int_0^t W'(s) ds$. Therefore W(t) is bounded. This implies the solution $(V(t), W(t))$ for $t\in [0, T)$ lies in a bounded set in $\mathbb{R}^n \times \mathbb{R}^{n\times n}$ which contradicts Theorem~\ref{thm:differential}.

Now $\lim_{t\rightarrow T} |V(t)|^2 = \lim_{t \rightarrow T} \sum_{j=1}^n \lambda_j^2(t) = + \infty$ implies, by Corollary~\ref{corollary:sol}(2), there exists $i$ for which $\lim_{t \rightarrow T} \lambda_i^2(t) = + \infty$. Furthermore, by Corollary~\ref{corollary:sol}(3), there exists an index $h$ for which $\lambda_h^2(t) \geq \lambda_j^2(t)$ near $T$ for all $j$. Thus $\lambda_h^2(t)$ has the largest growth rate tending $+ \infty$ as $t\rightarrow T$. 

For generic initial value $V(0)$ and $W(0)$, $\lambda_{h}^2(t)$ is the unique term of maximum growth rate. Therefore, we see that part (1) of Therem~\ref{thm:main_theorem} holds.

\medskip
\textbf{Case 2.} If $T = + \infty$, let us assume that the limit $\lim_{t \rightarrow T} |V(t)| = L > 0$ and show that structural balance occurs eventually.

By Corollary~\ref{corollary:sol}, we may assume that $\lambda_h^2(t) \geq \lambda_j^2(t)$ for $t$ near $\infty$ for all $j$  (if $\lim_{t \to T} |V(t)|^2=\infty$)  or  $\lim_{t \rightarrow T} \lambda_h^2(t) =\max\{ \lim_{ t \to T} \lambda^2(t) | i=1,..., n\} < \infty$.  Let $L' =\lim_{t \rightarrow T} \lambda_h^2(t) $. Then $L'>0$ since $L>0$. In the case of $\lim_{t \to T} |V(t)|^2=\infty < \infty$, for generic initial value, we may assume that $\lim_{t \rightarrow T} \lambda_h^2(t)$ is the unique maximum value among all $\lim_{t \rightarrow T} \lambda_i^2(t)$, $i=1, ..., n$.  Then we see that 
\begin{equation*}
    \int_0^t |\lambda_h^2(s)| ds \geq (t- t_0) L'' + c_0
\end{equation*}
for some constants $L''>0$ and $c_0$. It tends to $+ \infty$ as $t \rightarrow \infty$.

Furthermore, by the Cauchy inequality,
\begin{equation*}
    (\int_{0}^t \lambda_i(s) \lambda_j(s) ds )^2 \leq \int_{0}^{t} \lambda_i^2(s) ds \cdot \int_0^{t} \lambda_j^2(s) ds
\end{equation*}
and
\begin{equation*}
    \int_0^{t} \lambda_i^2(s) ds \leq \int_0^t \lambda_h^2(s) ds + c_1
\end{equation*}
for t large, we see the growth rate of 
$\int_{0}^t \lambda_i(s) \lambda_j(s) ds$ is at most that of $\int_0^t \lambda_h^2(s) ds$ as $t \to \infty$. 
This shows, by the same argument, that the growth rate of $w_{ij}(t)$ is dominated by $u_{ih} u_{jh}\int_0^t \lambda_h^2(s) ds$. Therefore, structural balance occurs again for generic initial values. 

Finally, we prove that in  Case 1 or in Case 2 that $T=\infty$ such that $\lim_{t \to \infty} |V(t)| =L >0$, the limit $\lim_{t \to \infty} V(t)/|V(t)|$ exists.

By corollary \ref{cor:v'} (2), $\lim_{t \to T} \lambda_i$ exists in $[-\infty, \infty]$. Therefore, if
 $\lim_{t \to \infty} |V(t)| =L$ is a finite positive number, then $\lim_{t \to \infty} V(t)$ exists in $\mathbb{R}^n-\{0\}$. Hence $\lim_{t \to \infty} V(t)/|V(t)|$ exists.  In the remaining cases, we have $\lim_{ t \to T} |V(t)|=\infty$. In this case, by the argument above, we see that $\lambda_i(t)$ has the same sign near $T$. We claim that the function
 $\lambda_i/\lambda_j$ is monotonic near $T$. Indeed, by the quotient rule for derivative,  the sign of derivative of $\lambda_i/\lambda_j$ is the same as that of $h(t)=\lambda_i'\lambda_j-\lambda_j'\lambda_i$. Now $h'(t) = \lambda_i''\lambda_j -\lambda_j''\lambda_i=
 (2|V|^2-a_i)\lambda_i\lambda_j -(2|V|^2-a_j)\lambda_j\lambda_i=(a_j-a_i)\lambda_i\lambda_j$.  Therefore, either $h'(t)$ has the same sign for $t$ near $T$ (when $a_i \neq a_j)$ or $h'(t)=0$ (when $a_i =a_j)$. If $h'(t)=0$, then $\lambda_i/\lambda_j$ is a constant near $T$ and the claim follows. If $h'(t)$ has the same sign near $T$, then $h(t)$ is strictly monotonic near $T$. Therefore,  $h(t)$ has the same sign for $t$ near $T$. As a consequence we see that $\lambda_i/\lambda_j$ is a monotonic near $T$. In particular, the limit $\lim_{ t \to T} \lambda_i(t)/\lambda_j(t)$ exists. Since $\lim_{t \to T} \sum_{k=1}^n \lambda_k(t)^2=\infty$,  this implies  the limit $\lim_{t \to T}\frac{\lambda_i(t)}{\sqrt{\sum_{j=1}^n \lambda_j(t)^2}}=\lim_{t \to T} \frac{\pm 1}{\sqrt{\sum_{j=1}^n \lambda_j(t)^2/\lambda_i(t)^2}}$ exists for any index $i$. The last statement is the same as that $\lim_{t \to T} V(t)/|V(t)|$ exists.

This ends the proof of Theorem~\ref{thm:main_theorem}.
\end{proof}


\begin{corollary}\label{cor:wv} Suppose $\lim_{t \to T} |V(t)|=\infty$. Then $A:=\lim_{ t \to T} W(t)^2/|V(t)|^2$ exists and $\lim_{t \to T} V/|V|$ is an eigenvector of $A$ associated to the eigenvalue one.
\end{corollary}

To see this, let $v$ be $\lim_{t \to T} V/|V|$. From $W^2=VV^T -C$ and $\lim_{t \to T} |V(t)| =\infty$, we see that
$W^2/|V|^2 =(V/|V|) (V/|V|)^T -C/|V|^2$ implies $\lim_{ t \to T} W^2/|V|^2 =vv^T$. Since $vv^Tv=v$ due to $|v|^2=1$, the result follows.

\medskip
We end the appendix by making several remarks and a conjecture.

The 1-dimensional case of equation $V''(t) = (2|V(t)|^2 - C)\cdot V(t)$, $V(t) \in \mathbb{R}^n$ is $y''(t) = 2y^3(t) - cy(t)$. The function $f(t) = \frac{a}{\sinh{(at + b)}}$ solves $f'' = 2f^3 + a^2f$ and $g(t) = \frac{a}{\sin(at+b)}$ solves $g'' = 2g^3 - a^2g$. We may assume $a>0$. Therefore, if $b<0$, then both $f$ and $g$ exist only on a finite maximum interval (it is $[0, -\frac{b}{a})$ for $f(t)$), i.e., $T < + \infty$.
If $b > 0$, then the function $f(t)$ exists on $[0, + \infty)$ but the function $g(t)$ exists only on a finite interval $[0, T)$. It indicates that if $C$ has a positive eigenvalue $a_1>0$, then the solution $\lambda_1(t)$ may exist only on a finite interval.  

This prompts us to conjecture that

\begin{conj}
 If the initial value matrix $C = V(0)V(0)^T - W(0)W(0)^T$ has a positive eigenvalue (i.e., $a_1 > 0$), then the maximum interval $[0, T)$ of existence for the solution $(V(t), W(t))$ of the co-evolution equation $V'=WV$ and $W'=VV^T$ is finite, i.e., $T < +\infty$.  
\end{conj}

If the conjecture holds, by Theorem~\ref{thm:main_theorem}, we see structural balance must occur eventually for generic initial value $C$ which has a positive eigenvalue.  Therefore, it also justifies our experimental observation that structural balance occurs almost all the time.

%% file: Riccati.tex
\section{Analysis of Social Tie Evolution}
\subsection{Structural Balance with $W(t)$}\label{subsec:structural}

\thmconvergence*
\begin{proof}
Let all eigenvalues of $W(t)$ be $\beta_1, ..., \beta_n$ and $\beta_1$ be the unique largest eigenvalue.  Then 
\begin{equation*}
    W(t) = H \cdot \diag(\beta_1(t), \cdots, \beta_n(t)) \cdot H^T,
\end{equation*}
for some time independent orthogonal matrix $H$ whose first column is the $\beta_1$ eigenvector. This implies that $w_{ij} = \sum_{k} H_{ik} \cdot \beta_k(t) \cdot H_{jk}$. Since all $\beta_i(t) < \beta_1(t)$ for $i\neq1$, the growth rate of $w_{ij}(t)$ as $t \to T$ is the same as $H_{i1}\beta_1(t) H_{j1}$. 

Therefore, the sign of $w_{ij}$ is the same as the sign of $H_{i1} \cdot H_{j1}$. Thus, 
\begin{equation*}
   \sgn( w_{ij}w_{jk}w_{ki}) = \sgn(H_{i1}H_{j1}H_{j1}H_{k1}H_{k1}H_{i1}) = \sgn(H_{i1}^2H_{j1}^2H_{k1}^2) >0.
\end{equation*}
The same argument shows that above sign is non-negative if some component of the eigenvector is zero. 
\end{proof}

\subsection{Solution to the General Riccati Equation}\label{subsec:general-Riccati}

We first focus on solving a general form of the matrix Riccati equation as stated below.
\begin{equation}
\label{eqn:Evolution}
    \left\{
    \begin{array}{rl}
        W' & = W^2 + C \\
        W(0) & = B.
    \end{array}
    \right.
\end{equation}
The equation in our co-evolution model, i.e.,  Equation~(\ref{eqn:Riccati}), satisfies $C+B^2=V(0)V(0)^T$. Notice that the right-hand side $V(0)V(0)^T$ is an $n
\times n$ matrix with rank one, which is a special condition. The analysis in this subsection applies for general matrices $B, C$.

By using a result in Reid~\cite{reid1972mathematics}, we can turn the matrix Riccati equation to a linear ODE system. 
\begin{lemma}[\cite{reid1972mathematics}]\label{lemma:recatti}
The ODE system in Equation~(\ref{eqn:Evolution}) is equivalent to the following system
\begin{equation}
\label{eqn:Rccati}
    \left\{
    \begin{array}{ll}
        Y' = Z  & Y(0) = I\\
        Z' = -CY & Z(0) = -B,
    \end{array}
    \right.
\end{equation}
where $Y' = -WY$ and $Z = -WY$, and $Y, Z, W \in \mathbb{R}^{n\times n}$.
\end{lemma}

\begin{proof}
First, we show that we can get Equations~(\ref{eqn:Rccati}) from Equations~(\ref{eqn:Evolution}), i.e., (\ref{eqn:Evolution}) $\Rightarrow$ (\ref{eqn:Rccati}).

We know that $Y$ exists and $Z = -WY$. Thus, $Y' = Z$. Then
\begin{equation*}
    \begin{split}
        Z' & = -W'Y - WY' \\
           & = -(W^2+C)Y - W(-WY) \\
           & = -W^2Y - CY + W^2Y \\
           & = -CY.
    \end{split}
\end{equation*}
At the same time, $Z(0) = -W(0)Y(0) =-W(0)= -B$. Thus, this direction is satisfied.

Second, let us prove that Equation~(\ref{eqn:Evolution}) can be obtained from  Equation~(\ref{eqn:Rccati}). 

Since $Y(0)=I$, by continuity,  $Y(t)^{-1}$ exists for $t \in [0, \epsilon]$, for $
\epsilon>0$. 
Since $Z=-WY$, we have $W = -ZY^{-1}$. 
Note $(Y^{-1})' = -Y^{-1}\cdot Y' \cdot Y^{-1}$, since $Y\cdot Y^{-1} = I \Rightarrow Y'\cdot Y^{-1} + Y\cdot (Y^{-1})' = 0$. 
Therefore:
\begin{equation*}
    \begin{split}
        W' & = -Z'\cdot Y^{-1} - Z \cdot (Y^{-1})' \\
           & = CY\cdot Y^{-1} - Z (-Y^{-1}\cdot Y' \cdot Y^{-1}) \\
           & = C + Z\cdot Y^{-1} \cdot (-WY) \cdot Y^{-1} \\
           & = C - Z\cdot Y^{-1} \cdot W \\
           & = C + W^2.
    \end{split}
\end{equation*}
Clearly $W(0) = -Z(0) = B$. This completes the proof.
\end{proof}

From Lemma~\ref{lemma:recatti}, we can focus on solving the linear Ordinary Differential Equation (ODE) in Equation~(\ref{eqn:Rccati}), which can be written in a matrix form. The analysis below is new. 
\begin{equation*}\label{eqn:matrix}
    \left[ \begin{array}{c}
Y \\
Z
\end{array}
\right]' = 
\left[ \begin{array}{cc}
0 & I \\
-C & 0
\end{array}
\right]
\left[ \begin{array}{c}
Y\\
Z
\end{array}
\right],
\end{equation*}
where we define $A = \left[ \begin{array}{cc}
0 & I \\
-C & 0
\end{array}
\right]\in \mathbb{R}^{2n \times 2n}$ in block form and $X = \left[ \begin{array}{c}
Y\\
Z
\end{array}
\right] \in \mathbb{R}^{2n \times n}$.

Now, let us solve the evolution equation $X' = AX$, where $X(0) = \left[ \begin{array}{c}
I\\
-B
\end{array}
\right]$. It is well known that the solution is,
\begin{equation}\label{eqn:X(t)}
    X(t) = (\sum_{n=0}^{\infty}\frac{t^n A^n}{n!}) X(0).
\end{equation}

Let us compute $A^n$ using the block multiplication of matrices~\cite{eves1980elementary,anton2013elementary}.

\begin{equation*}
    A^2 = \left[ \begin{array}{cc} 0 & I \\ -C & 0 \end{array} \right]
    \left[ \begin{array}{cc} 0 & I \\ -C & 0 \end{array} \right] =
    \left[ \begin{array}{cc} -C & 0 \\ 0 & -C \end{array} \right].
\end{equation*}
It implies:
\begin{equation*}
    A^{2n} = \left[ \begin{array}{cc} (-1)^n C^n & 0 \\ 0 & (-1)^n C^n \end{array}, \right]
\end{equation*}
\begin{equation*}
    A^{2n+1} = A^{2n}\cdot A = \left[ \begin{array}{cc} (-1)^n C^n & 0 \\ 0 & (-1)^n C^n \end{array} \right] \cdot
    \left[ \begin{array}{cc} 0 & I \\ -C & 0 \end{array} \right] = 
    \left[ \begin{array}{cc} 0 & (-1)^n C^{n} \\ (-1)^{n+1}C^{n+1} & 0 \end{array} \right].
\end{equation*}

Recall that $W = -ZY^{-1}$. Now we are ready to solve for $W$.

\mainth*
\begin{proof}

According to Equation (\ref{eqn:X(t)}),
\begin{equation*}
    \begin{split}
    X(t) & = \sum_{n=0}^{\infty}\frac{t^{2n}A^{2n}}{(2n)!}\cdot X(0) + \sum_{n=0}^{\infty}\frac{t^{2n+1}A^{2n+1}}{(2n+1)!}\cdot X(0) \\
         & = \sum_{n=0}^{\infty}\frac{(-1)^n t^{2n}}{(2n)!}\left[ \begin{array}{cc} C^n & 0 \\ 0 & C^n \end{array} \right] \left[ \begin{array}{c}1\\-B \end{array} \right] + \sum_{n=0}^{\infty}\frac{(-1)^n t^{2n+1}}{(2n+1)!} \left[ \begin{array}{cc}  0 & C^n \\ -C^{n+1} & 0 \end{array} \right] \left[ \begin{array}{c}1\\-B \end{array} \right] \\
         & = \sum_{n=0}^{\infty} \frac{(-1)^n t^{2n}}{(2n)!} \left[ \begin{array}{c}C^{n}\\-C^{n}B \end{array} \right] + \sum_{n=0}^{\infty}\frac{(-1)^{n+1} t^{2n+1}}{(2n+1)!} \left[ \begin{array}{c}C^{n}B\\C^{n+1} \end{array} \right] \\
         & = \left[ \begin{array}{c} \sum_{n=0}^{\infty} \frac{(-1)^n t^{2n} C^n}{(2n)!} + \sum_{n=0}^{\infty}\frac{(-1)^{n+1}t^{2n+1}C^nB}{(2n+1)!} \\
         \sum_{n=0}^{\infty}\frac{(-1)^{n+1}t^{2n}C^nB}{(2n)!} +  \sum_{n=0}^{\infty} \frac{(-1)^{n+1} t^{2n+1} C^{n+1}}{(2n+1)!} \end{array} \right] \\
         & \triangleq \left[ \begin{array}{c} Y \\Z  \end{array} \right].
    \end{split}     
\end{equation*}
\end{proof}

\subsection{When $B, C$ are both symmetric}\label{subsec:socialtie}

In our model, we assume that the initial tie matrix $W(0)$ is symmetric. Thus the social tie evolution follows $W' = W^2 + C$, where $W(0) = B$, both $B$ and $C$ are symmetric. We are able to derive more detailed closed form solutions for the matrix Riccati equation in this setting.

Since $C$ is symmetric, there exists an orthogonal matrix $U$ such that $U^TCU$ is a diagonal matrix:
$U^TCU=\diag\{a^2_1, \cdots, a^2_k, -d^2_1, \cdots, -d^2_l, 0, \cdots, 0\}$,
where $a_i>0, d_j>0$.  Furthermore, if $BC=CB$, by the simultaneous diagonalization theorem, we may choose $U$ such  that both $U^TCU$ and $U^TBU$ are diagonal. 
By Lemma~\ref{conjugation}(1), without loss of generality, we are going to solve the equation with the initial opinion vector $U^TV(0)$ and initial weight matrix $U^TW(0)U$. This leads to a system as below
\begin{equation*}
\label{eqn:evolution-diagonalized}
    \left\{
    \begin{array}{rl}
        (U^TWU)'  & = (U^TWU)^2 + U^T C U \\
        U^T W(0)U & = U^TB U.
    \end{array}
    \right.
\end{equation*}
The solution of this system can be easily transformed back to the solution to the original system by conjugation. For simplicity, the $(i,j)$-th entry of a matrix $M$ will be denoted by $M_{ij}$.  
Define $\beta_{ii}$ the $i$th diagonal element of $U^TBU$, i.e., $\beta_{ii}=(U^TBU)_{ii}$.

Let us now work out explicitly the matrices $Y$ and $Z$ in Theorem~\ref{mainth}.

\begin{enumerate}
    \item For the positive eigenvalue $a_i^2$ of $C$, 
\begin{equation*}
    \begin{split}
        &\sum_{n=0}^{\infty}\frac{(-1)^n t^{2n} a_i^{2n}}{(2n)!} = \cos{(a_i t)} \\
        &\sum_{n=0}^{\infty}\frac{(-1)^{n+1} t^{2n+1} a_i^{2n}}{(2n+1)!} = -\frac{1}{a_i}\sin{(a_i t)}.
    \end{split}
\end{equation*}
Thus, $(U^TYU)_{ii} = \cos{(a_i t)} - \frac{1}{a_i} \sin{(a_i t)}\cdot \beta_{ii}$. Similarly, we have
\begin{equation*}
    \begin{split}
        & \sum_{n=0}^{\infty}\frac{(-1)^{n+1} t^{2n} a_i^{2n}}{(2n)!} = -\cos{(a_i t)} \\
        & \sum_{n=0}^{\infty}\frac{(-1)^{n+1} t^{2n+1} a_i^{2n+2}}{(2n+1)!} = -a_i \sin{(a_i t)}.
    \end{split}
\end{equation*}
So, $(U^TZU)_{ii} = -\cos{(a_i t)}\beta_{ii} - a_i \sin{(a_i t)}$.
\item For the zero eigenvalues of $C$, i.e., $c_i = 0$, we have $(U^TYU)_{ii} = 1 - t\cdot \beta_{ii}$ and $(U^TZU)_{ii} = -\beta_{ii}$.

\item For the negative eigenvalue $c_i = -d_i^2$, we have
\begin{equation*}
    \begin{split}
        (U^TYU)_{ii} = \sum_{n=0}^{\infty} \frac{t^{2n}d_i^{2n}}{(2n)!} + (-1)\sum_{n=0}^{\infty}\frac{t^{2n+1} d_i^{2n}}{(2n+1)!}\beta_{ii} = \cosh{(d_i t)} - \frac{1}{d_i}\sinh{(d_i t)}\beta_{ii} \\
        (U^TZU)_{ii} = \sum_{n=0}^{\infty} (-1)\frac{t^{2n} d_i^{2n}}{(2n)!}\beta_{ii} + \sum_{n=0}^{\infty} \frac{t^{2n+1}d_i^{2n+2}}{(2n+1)!} = - \cosh{(d_i t)} \beta_{ii} + d_i \sinh{(d_i t)}.
    \end{split}
\end{equation*}
\end{enumerate}

So we can summarize the above formulas as: $U^TYU = D_1 - D_2 U^T BU$ and $U^TZU = D_3 - D_1 U^TBU$, where 
\begin{eqnarray*}
    D_1&=&\diag\{\cos{(a_1 t)}, \cdots,\cos{(a_k t)}, \cosh{(d_1 t)}, \cdots, \cosh{(d_l t)}, 1, \cdots, 1 \}, \\
    D_2&=&\diag\{\sin(a_1 t)/a_1, \cdots,\sin(a_kt)/a_k, \sinh(d_1 t)/d_1, \cdots, \sinh(d_l t)/d_l, 1, \cdots, 1 \},\\
    D_3&=&\diag\{-a_1\sin(a_1 t), \cdots,-a_k\sin(a_kt), d_1\sinh(d_1 t), \cdots, d_l\sinh(d_l t), 0, \cdots, 0 \},
\end{eqnarray*} 

\subsection{When both $B, C$ are symmetric and $BC=CB$}\label{subsec:BCequalCB}

Now, we consider the special case that $BC=CB$.
Using a basic fact that two commuting symmetric matrices can be simultaneously orthogonally diagonalized~\cite{Hoffman1971-bc}. Let $U$ be an orthogonal matrix such that 
\begin{equation*}
    \begin{split}
        U^TCU & = \diag(a_1^2, \cdots, a_k^2, -d_1^2, \cdots, -d_l^2, 0, \cdots, 0)\\
        U^TBU & = \diag(\lambda_1, \cdots, \lambda_k, \mu_1, \cdots, \mu_l, \delta_1, \cdots, \delta_h).
    \end{split}
\end{equation*}
where $a_i, d_j > 0$.

In this case we can further simplify the solution.  
\begin{equation}\label{eqn:WWW}
\begin{aligned}
    W  = & -Z \cdot Y^{-1} \\
       = & U\diag(\frac{a_1 \sin{(a_1 t)} + \cos{(a_1 t)}\lambda_1}{\cos{(a_1 t)}-\frac{1}{a_1}\sin{(a_1 t)}\lambda_1}, \cdots, \frac{a_k \sin{(a_k t)}) + \cos{(a_k t)}\lambda_k}{\cos{(a_k t)}-\frac{1}{a_k}\sin{(a_k t)}\lambda_k},\\
       & -\frac{d_1 \sinh{(d_1 t)}-\cosh{(d_1 t)}\mu_1}{\cosh{(d_1 t)} - \frac{1}{d_1}\sinh{(d_1 t)}\mu_1}, \cdots, -\frac{d_l \sinh{(d_l t)}-\cosh{(d_l t)}\mu_l}{\cosh{(d_l t)} - \frac{1}{d_l}\sinh{(d_l t)}\mu_l}, \\
       & \frac{\delta_1}{1 - t\delta_1}, \cdots, \frac{\delta_h}{1 - t\delta_h})U^T.
\end{aligned}
\end{equation}

Note that above equation for $W(t)$ implies that $V(t)$ is an eigenvector of $W(t)$ for all time $t$.

Now we are ready to analyze the behavior of $W$ over time in the case of a symmetric initial condition matrix $W(0)$. 
We start with a technical lemma.
\begin{lemma}\label{lemma:sin}
\begin{enumerate}
\item If $a > 0$, then there exists $T \in (0, \frac{\pi}{a})$, such that for all $\lambda$
\begin{equation*}
    \lim_{t \rightarrow T} \frac{a \sin{(at)} + \lambda \cos(at)}{-\frac{\lambda}{a}\sin{(at)}+ \cos(at)} = + \infty.
\end{equation*}

\item If $d > 0$, then there exists $T \in (0, \infty]$, such that
\begin{equation*}
    \lim_{t \rightarrow T} \frac{\cosh(dt)\mu-d\sinh(dt)}{\cosh(dt)-\frac{\mu}{d}\sinh(dt)} = + \infty,
\end{equation*}
if and only if $\mu >d$. In the case $\mu \leq d$, the limit  for $T=\infty$ exists and is finite. 

\item If $\delta > 0$, then there exists $T = \frac{1}{\delta}$, such that 
$$\lim_{t \rightarrow T} \frac{\delta}{1 - t\delta} = + \infty.$$
\end{enumerate}
For all cases mentioned above, the convergence rate is $O(\frac{1}{|T-t|})$.
\end{lemma}

\begin{proof} For (1), we reorganize 
$$\frac{a \sin{(at)} + \lambda \cos(at)}{-\frac{\lambda}{a}\sin{(at)}+ \cos(at)} =\lambda\frac{\cot{(at)} + \frac{a}{\lambda}}{\cot{(at)}-\frac{\lambda}{a}},$$
Since $\cot{(at)}$ is strictly decreasing from $\infty$ to $-\infty$ in the range $(0, \frac{\pi}{a})$, there exists a $T \in (0,\frac{\pi}{a})$, such that $\cot{(at)} = \frac{\lambda}{a}$. As $t$ approaches $T$ from the left, $\cot{(at)} - \frac{\lambda}{a} > 0$. 

When $t \rightarrow T$ the numerator becomes $$\lambda(\frac{\lambda}{a} + \frac{a}{\lambda}) = \frac{\lambda^2 + a^2}{a}>0.$$
This confirms (i). 

Then we consider the convergence rate as $t \rightarrow T$. 
\begin{equation*}
        \lambda\frac{\cot{(at)} + \frac{a}{\lambda}}{\cot{(at)}-\frac{\lambda}{a}}  = \lambda \frac{\cot{(-a(T - t) + aT)} + \frac{a}{\lambda}}{\cot{(-a(T - t) + aT)} - \frac{\lambda}{a}}  = \lambda \frac{(\frac{\lambda}{a}+\frac{a}{\lambda})\cot{(a(T-t))}}{1 + \frac{\lambda^2}{a^2}}.
\end{equation*}
The Taylor series of $\cot{(x)}$ is $\frac{1}{x} - \frac{x}{3} + o(x)$. Given that $(T-t) \rightarrow 0$, the convergence rate is $O(\frac{1}{|T-t|})$. 

For (2), reorganize 
$$\frac{\cosh(dt)\mu-d\sinh(dt)}{\cosh(dt)-\frac{\mu}{d}\sinh(dt)}=\frac{\coth(dt)\mu-d}{\coth(dt)-\frac{\mu}{d}}.$$
Since the function $\coth(dt)$ is strictly decreasing from $\infty$ to $1$, if $\mu>d$, there exists $T \in (0, \infty)$ such that $\coth(dT)=\mu/d$. Furthermore, as $t$ approaches $T$ from the left, the denominator is positive. But $\lim_{t \to T}\coth(dt)\mu-d = \frac{\mu^2-d^2}{d} >0  $.
It follows that $$\lim_{t \rightarrow T} \frac{\cosh(dt)\mu-d\sinh(dt)}{\cosh(dt)-\frac{\mu}{d}\sinh(dt)} = + \infty.$$  If $\mu \leq d$, the function is smooth on $[0, \infty)$ and the limit is finite as $t \to \infty$. 

For convergence rate, we rewrite the function:
\begin{equation*}
        \mu \frac{\coth{(dt)} - \frac{d}{\mu}}{\coth{(dt)} - \frac{\mu}{d}}  = 
        \mu \frac{\coth{(-d(T - t) + dT)} - \frac{d}{\mu}}{\coth{(-d(T - t) + dT)} - \frac{\mu}{d}} = \mu \frac{(\frac{\mu}{d} - \frac{d}{\mu})\coth{(d(T-t))}}{\frac{\mu^2}{d^2} - 1}.
\end{equation*}
The Taylor series of $\coth{(x)}$ is $\frac{1}{x} + \frac{x}{3} - o(x)$. Given that $(T-t) \rightarrow 0$, the convergence rate is $O(\frac{1}{(T-t)})$.

The limit in case (3) is obvious when $\delta > 0$. Now we look at its convergence rate. Here $T = \frac{1}{\delta}$. Rewrite the function as 
\begin{equation*}
    \frac{\delta}{1 - t\delta} = \frac{1}{\frac{1}{\delta} - t} = \frac{1}{T - t} = \frac{1}{(T-t)}.
\end{equation*}
Thus, its convergence is $O(\frac{1}{(T-t)})$. 

From the above analysis, we can know the convergence rate is $O(\frac{1}{(T-t)})$, which is an inverse proportional function, under any case.
\end{proof}

By Equation~(\ref{eqn:W}) and Lemma~\ref{lemma:sin}, we know that some  diagonal entry of $W$ converges to the infinity at a finite time $T$ under some appropriate conditions.

\subsection{Structural Balance when $BC=CB$}\label{subsec:structural-commute}

In the following, we prove Theorem~\ref{thm:main} which characterizes the conditions for structural balance, when $BC=CB$.

\thmmain*
\begin{proof} Since the eigenvalues of $W(t)$ remains the same as $U^TWU$ for any orthogonal matrix $U$, we see the convergence of eigenvalues of $W(t)$ from Equation~(\ref{eqn:W}). Due to the convergence of the eigenvalues in $(-\infty, \infty]$, we see that  $\lim_{t \to T} w_{ij}(t) \in [-\infty, \infty]$ exists for all $i,j$. This means that we have sign stability, that all weights $w_{ij}$, $\forall i, j$, have fixed signs, as $t \to T$.

To see structure balance, by Theorem~\ref{kl}, it suffices to check if the largest eigenvalue tends to infinity. We examine the solution $W$ as described in Equation~(\ref{eqn:W}) and use Lemma~\ref{lemma:sin}.
\begin{enumerate}
\item If there exists $\delta_i > 0$, structural balance occurs because $\lim_{t\rightarrow \frac{1}{\delta_i}}\frac{\delta}{1-t \delta_i} = +\infty$.

\item If there exists $a_i>0$, structural balance occurs because of Lemma~\ref{lemma:sin}(1). 

\item If $\mu_j > d_j$, structural balance occurs because of Lemma~\ref{lemma:sin} (2).
\end{enumerate}
\end{proof}

Now we are ready to discuss our co-evolution model (Equation~(\ref{eqn:continuous-model})). First, the condition $BC=CB$ when $B=W(0)$ and $C=V(0)V(0)^T-W(0)W(0)^T$ means that $V(0)$ is an eigenvector of $W$. We finish the proof here. 

\lemmaBCcommute*
\begin{proof} Clearly if $Av=\alpha v$, then $A$ and $vv^T$ commute.   Conversely, if $A$ and $vv^T$ commute, we can find an orthogonal matrix $U$ such that $U^TAU$
and $U^Tvv^TU$ are diagonal. We may assume that the $(1,1)$ entry 
$\lambda$ of 
$U^Tvv^TU$ is not zero. This shows the first column $c$ of $U$ is an eigenvector for  $vv^T$ associated to $\lambda$.
But $v$ is also an eigenvector of $vv^T$ associate to $\lambda$. Therefore $c$ is a
non-zero scalar multiplication of $v$.  But we also know that $c$ is an eigenvector of $A$.  Therefore, $v$ is an eigenvector of $A$.
\end{proof}

For Equation~(\ref{eqn:continuous-model}), we have an additional condition $B^2+C = V(0)V(0)^T$ which has rank one. In the discussion below, we need the following fact about the eigenvalues and eigenvectors of rank-1 symmetric matrices $H=uu^T$ where $u \in \mathbb {R}^n -\{0\}$.  The matrix $H$ has eigenvalues $||u||^2$ and $0$ such that the associated eigenvectors are $u$ and $z$'s which are perpendicular to $u$, i.e., $u^Tz=0$. Therefore, the unique largest positive eigenvalue is $||u||^2$ with the associated eigenvector $u$. 

\corcomute*
\begin{proof}
When $V(0)$ is not a zero vector, $V(0)V(0)^T$ is a symmetric matrix with one positive eigenvalue and $(n-1)$ zero eigenvalue. Furthermore, $V(0)$ is an eigenvector associated to the largest positive eigenvalue.  Now, since $W(0)$ and $C$ commute, we may simultaneously orthogonally diagonalize both. Since $B^2+C=V(0)V(0)^T$ is diagonal with only one positive diagonal entry,  using the same notation as above, we see that all numbers $a_1^2+\lambda_1^2, ...., a_k^2+\lambda_k^2, \mu_1^2-d_1^2, ..., \mu_l^2-d_l^2, \delta_1^2, ..., \delta_n^2$ are zero except one of them which is positive. 
If $a_i^2+\lambda_i^2 > 0$, then $a_i > 0$ exists and the condition (2) holds. If $\mu_i^2 - d_i^2 > 0$, then $\mu_i > d_i$ exists and the condition (3) holds. If $\delta_i^2 + 0 > 0$, then $\delta_i > 0$ and the condition (1) holds. Thus, by Theorem~\ref{thm:main},  the structural balance must occur in the finite time for $W(t)$.
\end{proof}





%% file: extension.tex
\subsection{Extensions and Conjectures}
\label{subsec:general}

\noindent\textbf{General Graphs} The discussion so far has assumed a complete graph. For a connected graph $G$, with our evolution model, our conjecture is that all the weights and opinions will go to extreme values. This appears to be the case for all simulations we have run. The reason is that it is hard for the weights to converge to finite values. One exception is that all the opinions are $0$. But this is not a stable state. Any small perturbation on the opinion will break this stable state.

At convergence, however, there might be more than $2$ communities. The reason is that the edges with negative weights may not be in any triangle of the graph. Thus, multiple communities can be separated by the negative edges and it does not break the structure balance requirement.

\smallskip\noindent\textbf{High Dimensional Opinions} When each node has multiple opinions on different issues, its opinion can be represented as a $m$-dimensional vector, where $m$ is the number of opinions in one node. Each entry of the weight matrix is a $m\times m$ matrix instead of a real number. 

For the high dimensional setting,  our 
equation for a general graph is:
\begin{equation}\label{eqn:derivation}
    \left\{
    \begin{array}{l}
        V_i' = \sum_{j \sim i}W_{ij}V_j \\
        W_{ij}' = V_i V_j^T, i \sim j
    \end{array}
    \right.
\end{equation}
where $i\sim j$ means that there exists an edge between nodes $i$ and $j$. Then, we have the following theorem:
\begin{theorem} If the graph has no self-loop and consider tie matrices $W$ to be symmetric, i.e., $W=W^T$, then 
 Equation~(\ref{eqn:derivation}) is the gradient flow of the dissonance function:
\begin{equation*}
    F(V, W) = \frac{1}{2} \sum_{i\sim j} V_i^T \cdot W_{ij} \cdot V_j.
\end{equation*}
\end{theorem}
\begin{proof}
Since each opinion vector is $m$-dimensional, we can write it as $V_i = [(v_i)_a]\in \mathbb{R}^m$. Similarly, the edge weight $W_{ij}$ can be written as an $m\times m$ matrix $ [(w_{ij})_{ab}]\in \mathbb{R}^{m\times m}$. Note that $W = W^T$ means $(w_{ij})_{ab} = (w_{ji})_{ba}$.

 Equation~(\ref{eqn:derivation}) is the same as:
\begin{equation*}\label{eqn:derivation_detail}
    \left\{
    \begin{array}{l}
        (v_i)_a' = \sum_{j \sim i} \sum_{b=1}^m (w_{ij})_{ab} (v_j)_b \\
        (w_{ij})_{ab}' = (v_i)_a (v_j)_b, i \sim j.
    \end{array}
    \right.
\end{equation*} 
Due to $i \sim j$ implies $i \neq j$, the derivative of $F$ are:
\begin{equation*}
    \frac{\partial F}{\partial (w_{ij})_{ab}} = \frac{1}{2}(v_i)_a(v_j)_b+\frac{1}{2} (v_j)_b(v_i)_a  \triangleq (w_{ij})_{ab}'.
\end{equation*}
\begin{equation*}
\begin{split}
    \frac{\partial F}{\partial (v_i)_a} & = \frac{1}{2} \sum_{j \sim i} (v_j)_b(w_{ji})_{ba} + \frac{1}{2} \sum_{ k\sim i} (w_{ik})_{ac}(v_k)_c \\
    & = \sum_{j\sim i} (w_{ij})_{ab}(v_j)_b \triangleq (v_i)_a'.
\end{split}
\end{equation*}
\end{proof}

Based on our evolution equation and previous properties in the $1$-dimension opinion case, we make the following conjecture:

For a complete graph with self-loop edges, all the opinion vectors and the weight matrix will converge to extreme values. Any two adjacent nodes $V_i(t)$ and $V_j(t)$ have the same opinion or the complete opposite opinion, i.e., $V_i(t) = V_j(t)$ or $V_i(t) = -V_j(t)$ as time approaches $T_0$.

Based on the above conjecture, each entry of $\lim_{t \to T_0} W_{ij}(t)$ should have the same sign as that of  $\lim_{t \to T_0} V_i(t)V_j(t)^T$. 

%% file: simulation.tex
\section{Simulation}
We use the discrete time model as described in Equation~(\ref{eq:discrete-model}). The convergence of the dynamic process defined by Equation~(\ref{eq:discrete-model}) is very fast. For visualization purposes, in the simulation, we set our evolution model using Equation~(\ref{modii}) with $a = b = 0.01$. 
When a graph is said to satisfy structural balance, the multiplication of weights along all triangles is non-negative. 
In addition, either all nodes have the same opinion (i.e., harmony) or the graph is partitioned into two antagonistic groups (i.e., polarization).

\subsection{Harmony vs Polarization}
In this section, we work with a complete graph and examine when structural balance and/or polarization appears. 

\begin{figure}[tb]
\centering
\subfigure[$W(0) = I$]{
\centering
\includegraphics[width=0.23\columnwidth]{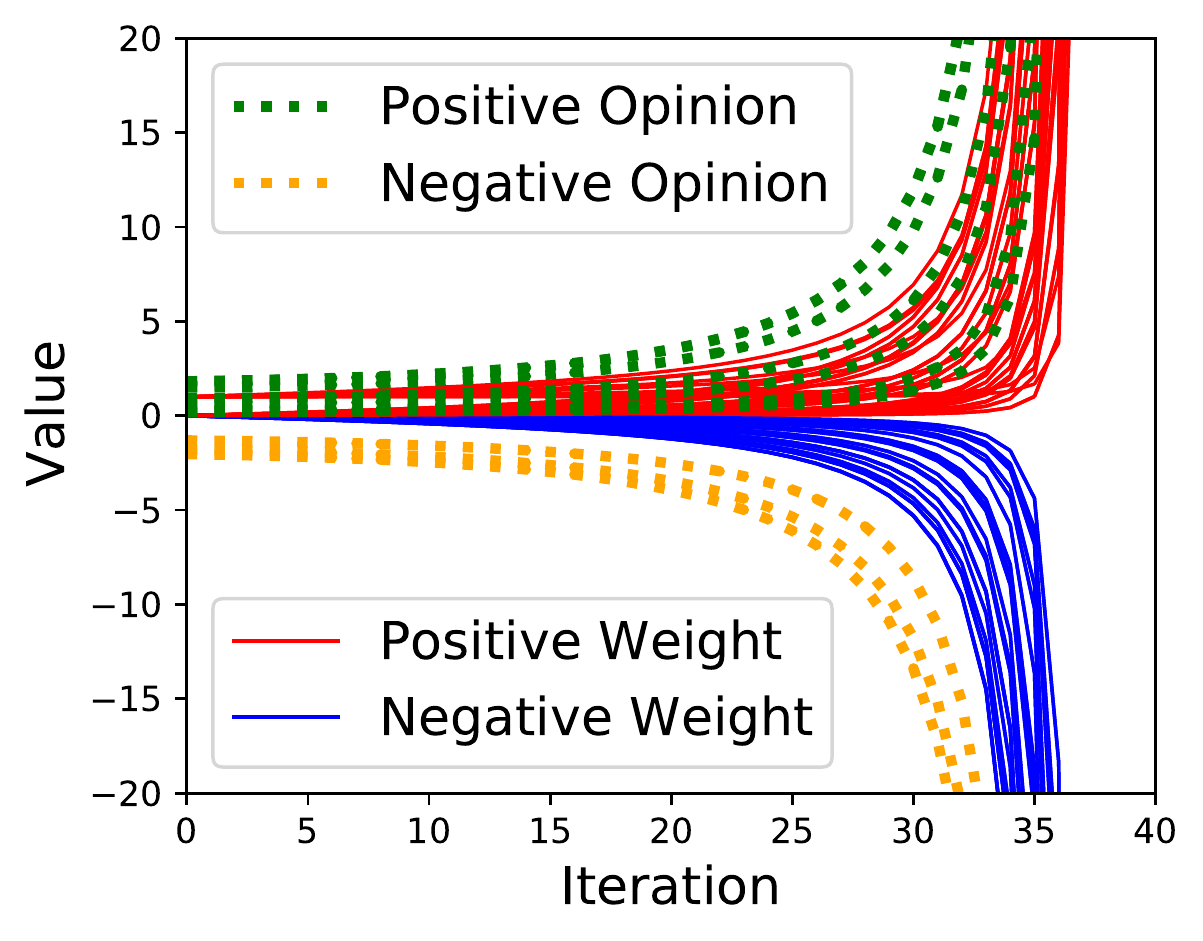}
\label{fig:case0_P}
}
\subfigure[$W(0) = 0$]{
\centering
\includegraphics[width=0.23\columnwidth]{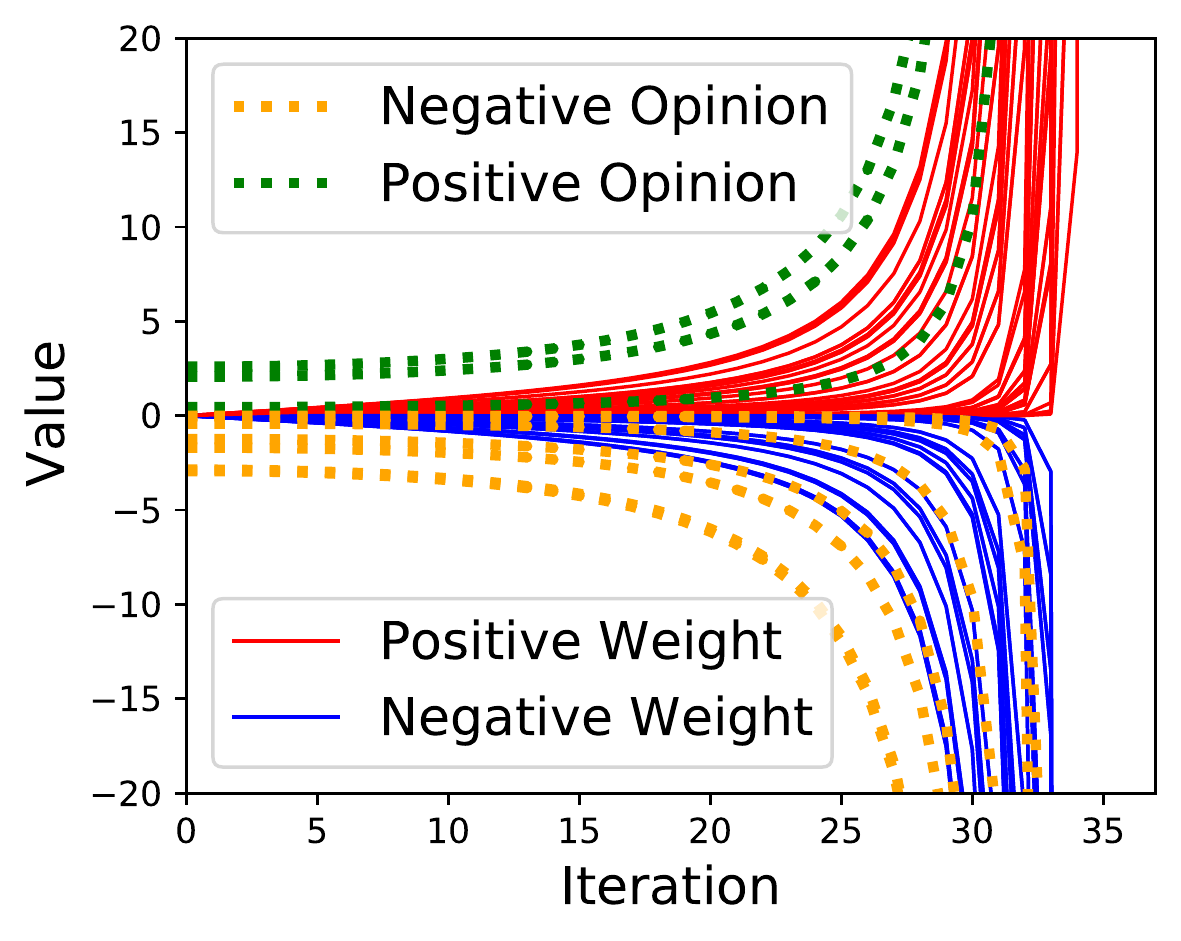}
\label{fig:case1_P}
}
\subfigure[$W(0) = V(0)V(0)^T$]{
\centering
\includegraphics[width=0.23\columnwidth]{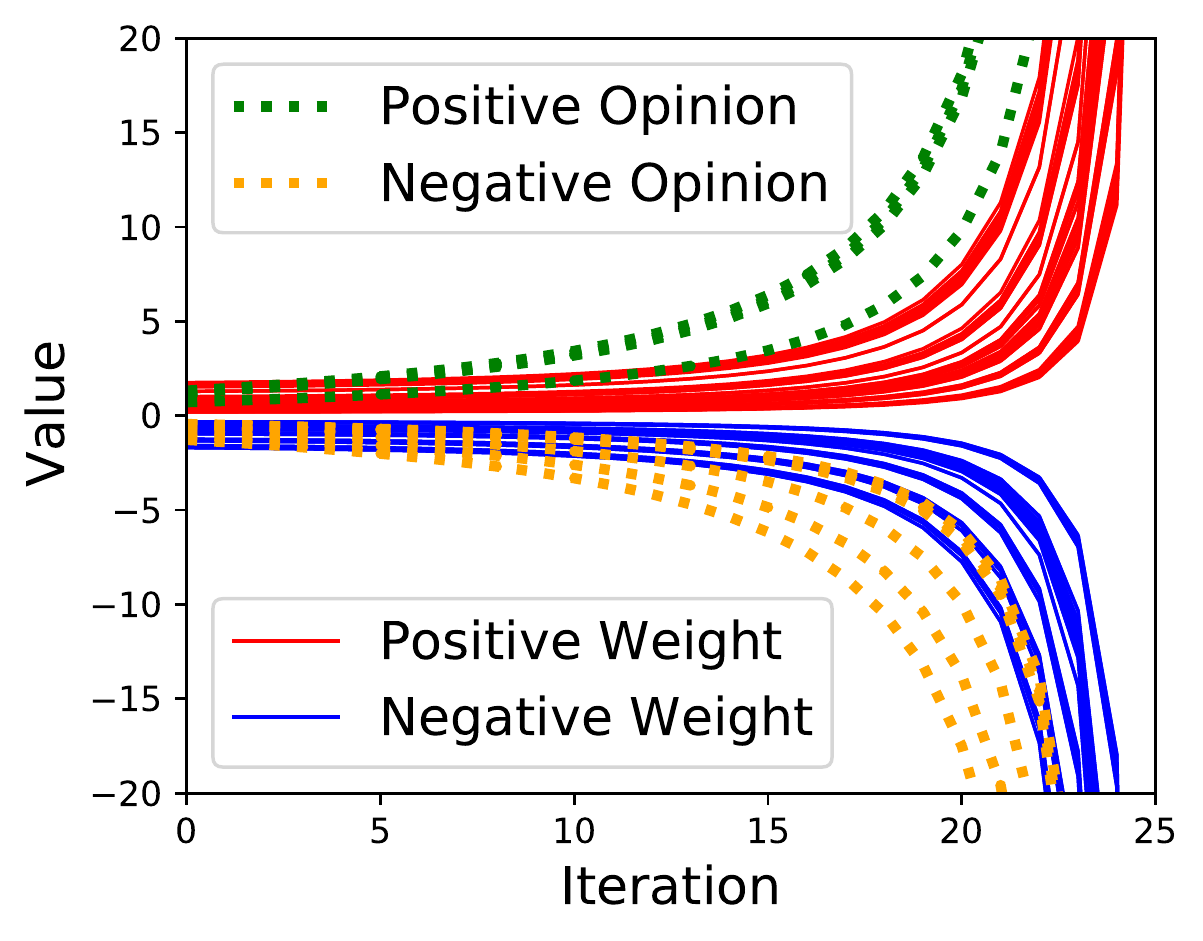}
\label{fig:case2_P}
}
\subfigure[$W(0) = -2I$]{
\centering
\includegraphics[width=0.23\columnwidth]{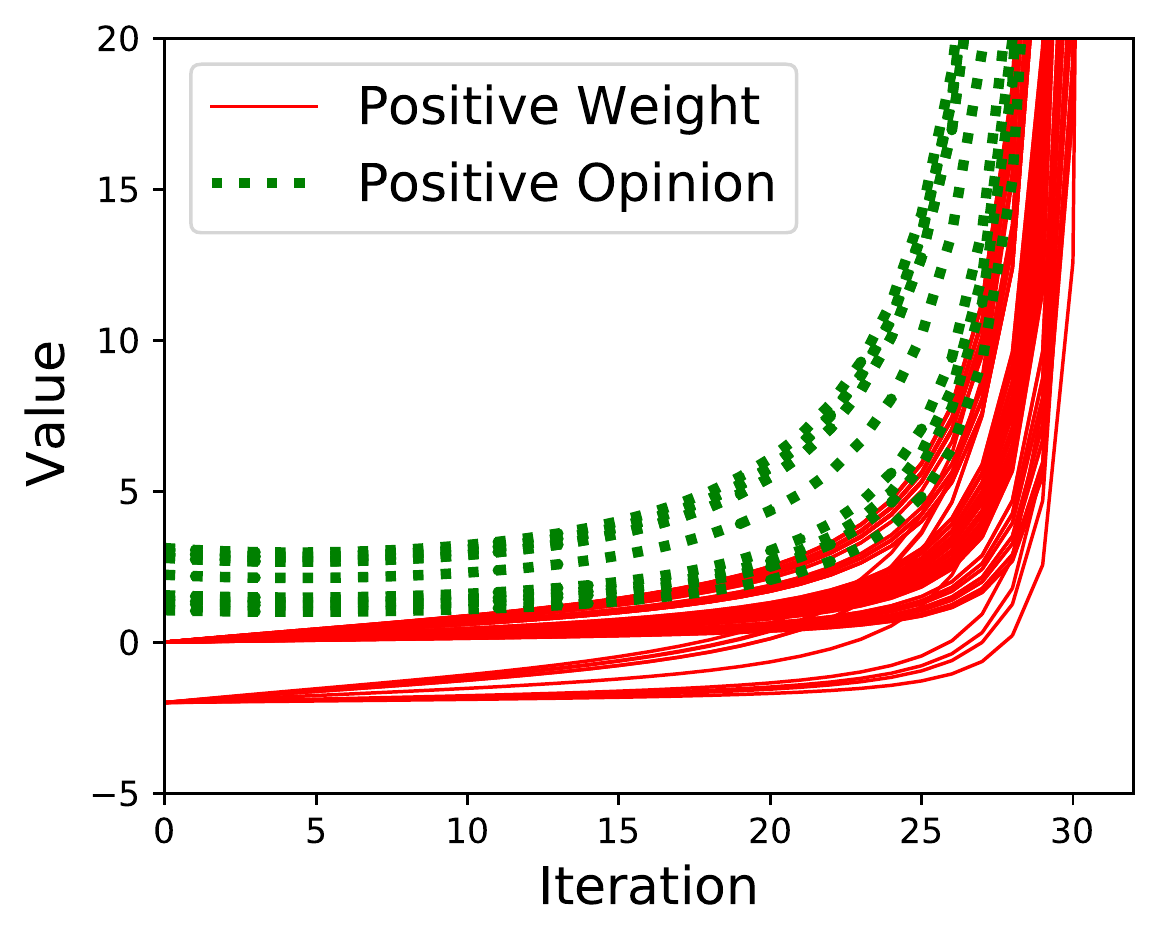}
\label{fig:case0_C}
}
\caption{The evolution process for a complete graph with $BC=CB$. In the first three plots, the network at the limit achieves structural balance and the network is also polarized. In the last example, the network reaches harmony. }
\label{fig:commute_evolution}
\end{figure}

\smallskip\noindent\textbf{$B$ and $C$ commute. } As mentioned in Theorem~\ref{thm:main}, when the matrices $B = W(0)$ and $C = V(0)V(0)^T - W(0)^2$ are symmetric with $BC=CB$, structural balance occurs in finite time for $W(t)$ if $W(t)$ has an unique largest eigenvalue. We take three special cases mentioned in the previous section, namely when $W(0) = I$, $W(0) = 0$, or $W(0) = V(0)V(0)^T$. The evolution process of these three cases are shown in Figure~\ref{fig:commute_evolution}. The initial opinions are selected uniformly at random in $[-1, 1]$. The dash curves represent the evolution of opinions and the solid ones shows the edge weights. The color of the curves represent the final sign of opinions and weights at convergence. In all three cases the network reaches polarization where the opinions of some vertices go to positive infinity and the opinion of the others go to negative infinity. In these cases, the node opinions do not change signs, because $v_iw_{ij}v_j \geq 0$ holds all the time. The gradient direction of opinions and weights are same with their signs. 
Figure~\ref{fig:case0_C} shows a case when both opinions and edge weights converge to positive values. All the opinions are assigned initially as a positive value. The initial weight matrix is a diagonal matrix with the same diagonal entry. All final opinions and weights are positive. 

\begin{figure}[tb]
\centering
\subfigure[Symmetric matrix $W(0)$]{
\centering
\includegraphics[width=0.23\columnwidth]{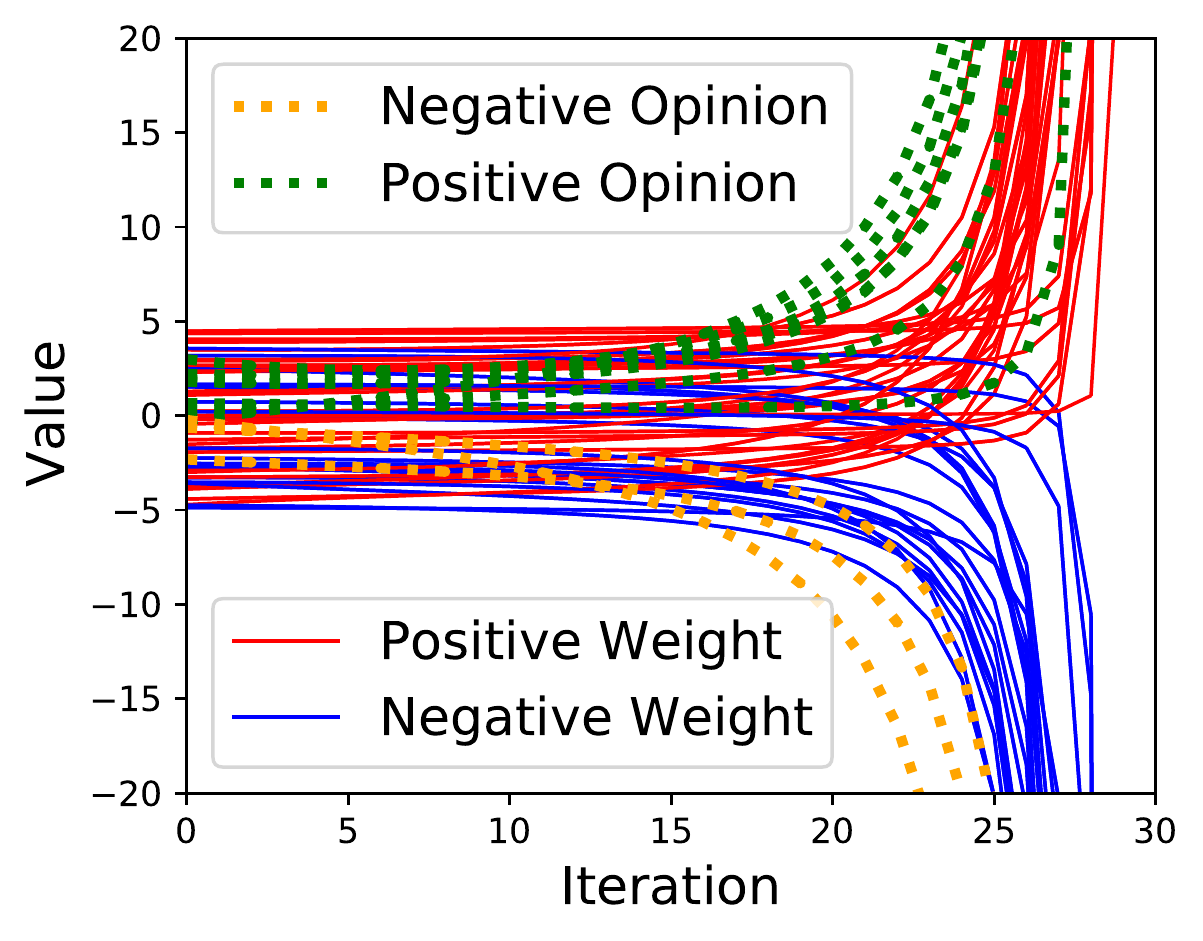}
\label{fig:case0_NC}
}
\subfigure[Symmetric matrix $W(0)$]{
\centering
\includegraphics[width=0.23\columnwidth]{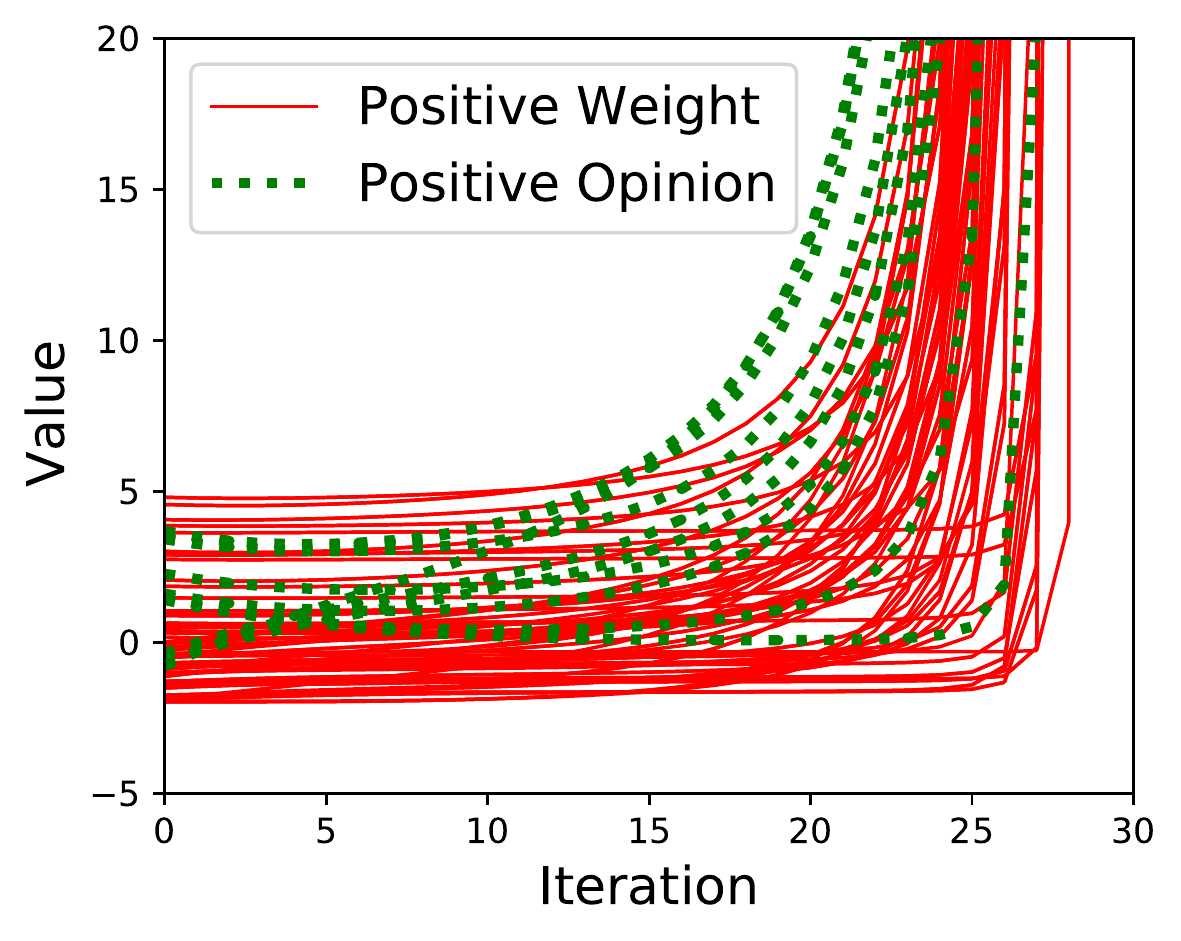}
\label{fig:case1_NC}
}
\subfigure[Random matrix $W(0)$]{
\centering
\includegraphics[width=0.23\columnwidth]{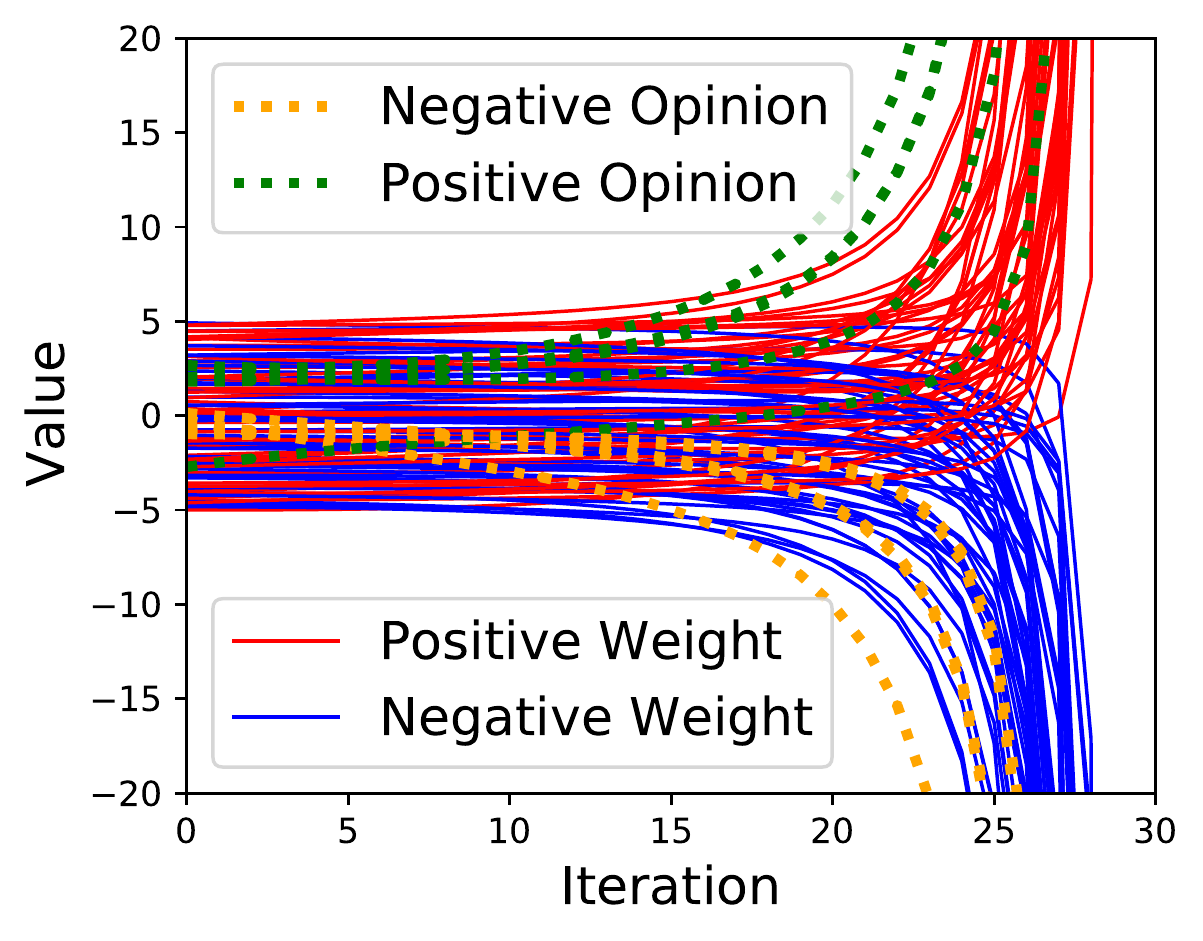}
\label{fig:case2_NC}
}
\subfigure[Negative matrix $W(0)$]{
\centering
\includegraphics[width=0.23\columnwidth]{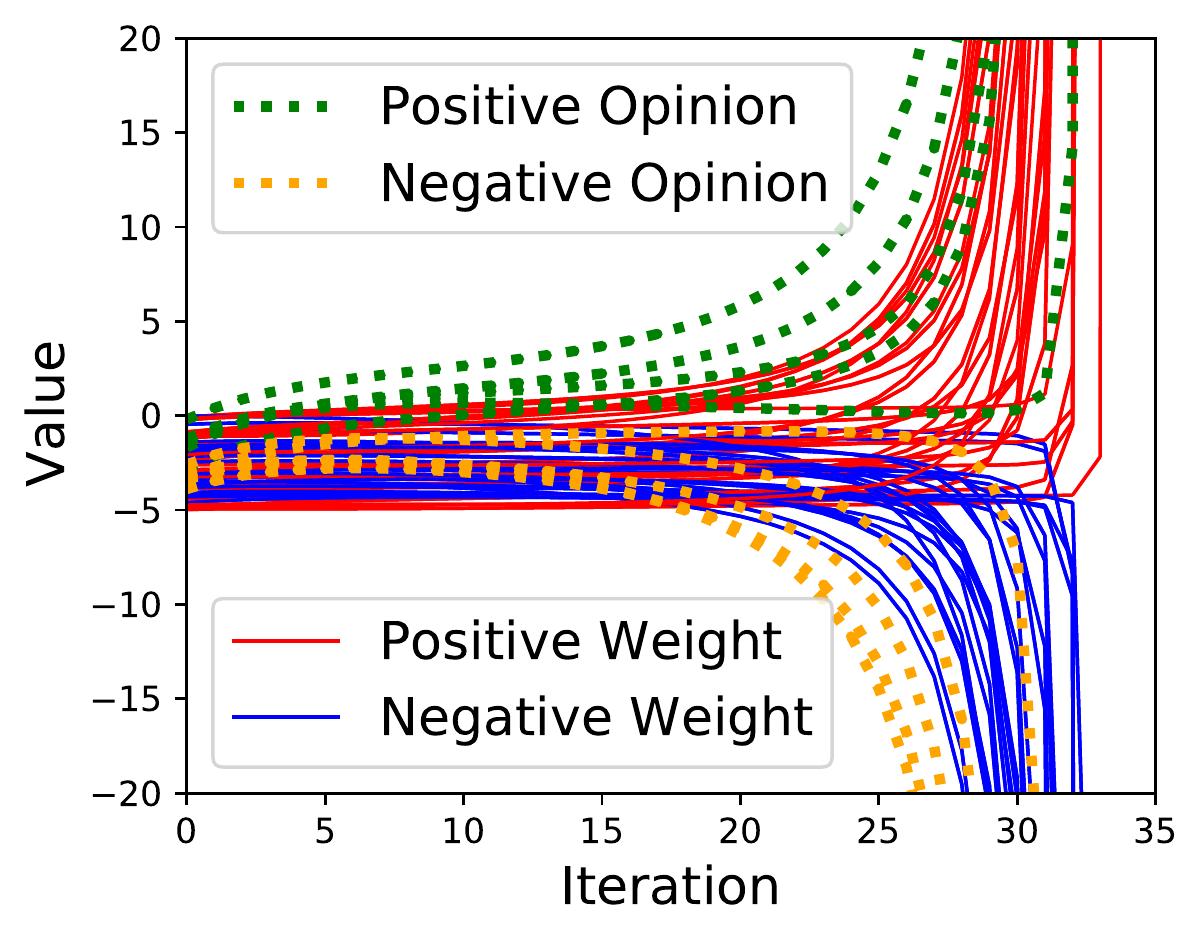}
\label{fig:case3_NC}
}
\caption{The evolution process for a complete graph with $BC\neq CB$. In both (a) and (b) we take a symmetric random matrix as $W(0)$ and a random initial vector $V(0)$. In (a) the network reaches a polarized state and in (b) the network reaches harmony. Figure~\ref{fig:non_commute_evolution}(c) shows the evolution when the initial weight matrix is a random non-symmetric matrix. Some edge weights and opinions change signs during the process. In Figure~\ref{fig:non_commute_evolution}(d), all the entries in the initial matrix and the opinion vector are negative. The network reaches a polarized state and reaches structural balance.}
\label{fig:non_commute_evolution}
\end{figure}

\smallskip\noindent\textbf{$B$ and $C$ do not commute. } Next, we consider the case when the matrices $B$ and $C$ do not commute. We take different random  initial cases and show the evolution process in Figure~\ref{fig:non_commute_evolution}. In the first two plots ~\ref{fig:case0_NC} and ~\ref{fig:case1_NC}, we select a symmetric random matrix as the initial weight matrix $W(0)$. The opinions are assigned random initial values. In simulation we observed both cases of harmony and polarization as the final state. Some edges and vertices change signs in the process. We also show an example of the evolution process in Figure~\ref{fig:heatmap} which shows how two communities emerge. 

\smallskip\noindent\textbf{$W(0)$ is not symmetric. } In Figure~\ref{fig:case2_NC}, the initial weight matrix is not symmetric. Both vertex opinions and edge weights go to infinity after several iterations. Figure~\ref{fig:case3_NC} shows when all the entries in the matrix and opinions are initially negative. Some opinions and weights change signs in the evolution process. For all cases we have tested when $B$ and $C$ do not commute, structural balance is always satisfied in the limit. 

\subsection{Convergence Rate}
In section, we check the convergence rate in different settings. This helps us understand intuitively the factors that influence the convergence rate. 

\begin{figure}[tb]
\centering
\subfigure[$W(0) = I$]{
\centering
\includegraphics[width=0.3\columnwidth]{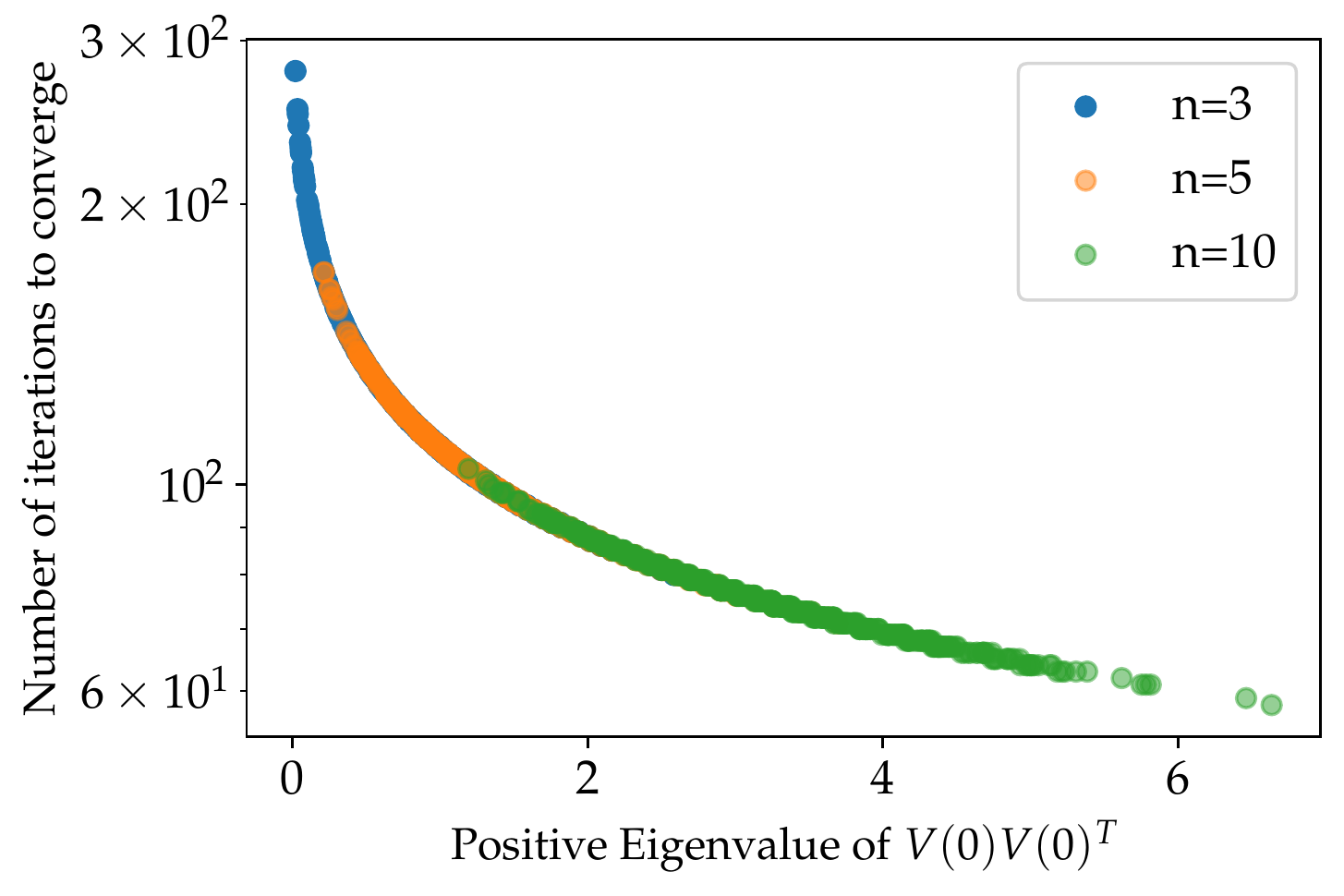}
\label{fig:convergence_1}
}
\subfigure[$W(0) = 0$]{
\centering
\includegraphics[width=0.28\columnwidth]{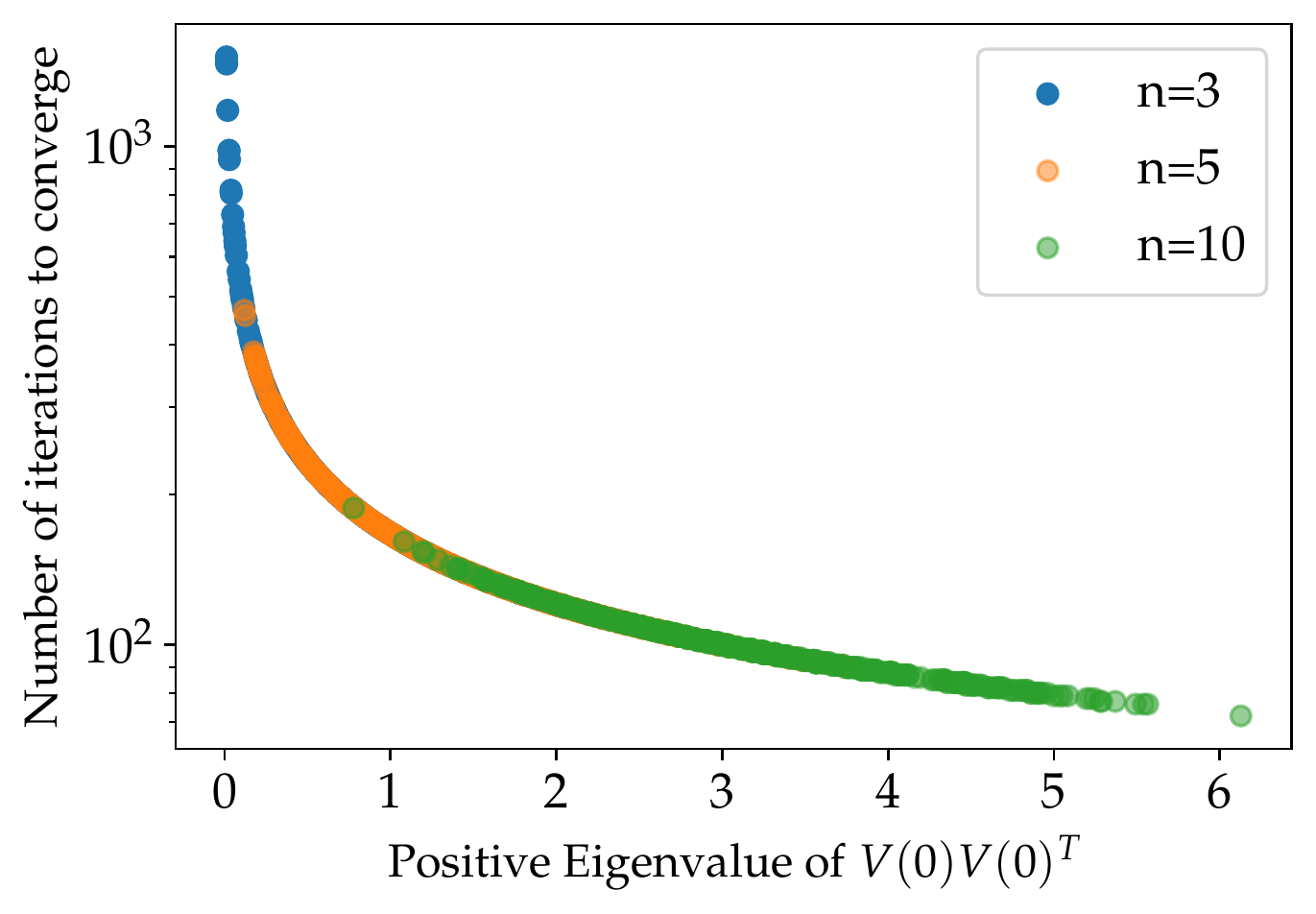}
\label{fig:convergence_2}
}
\subfigure[Non-commutative cases]{
\centering
\includegraphics[width=0.3\columnwidth]{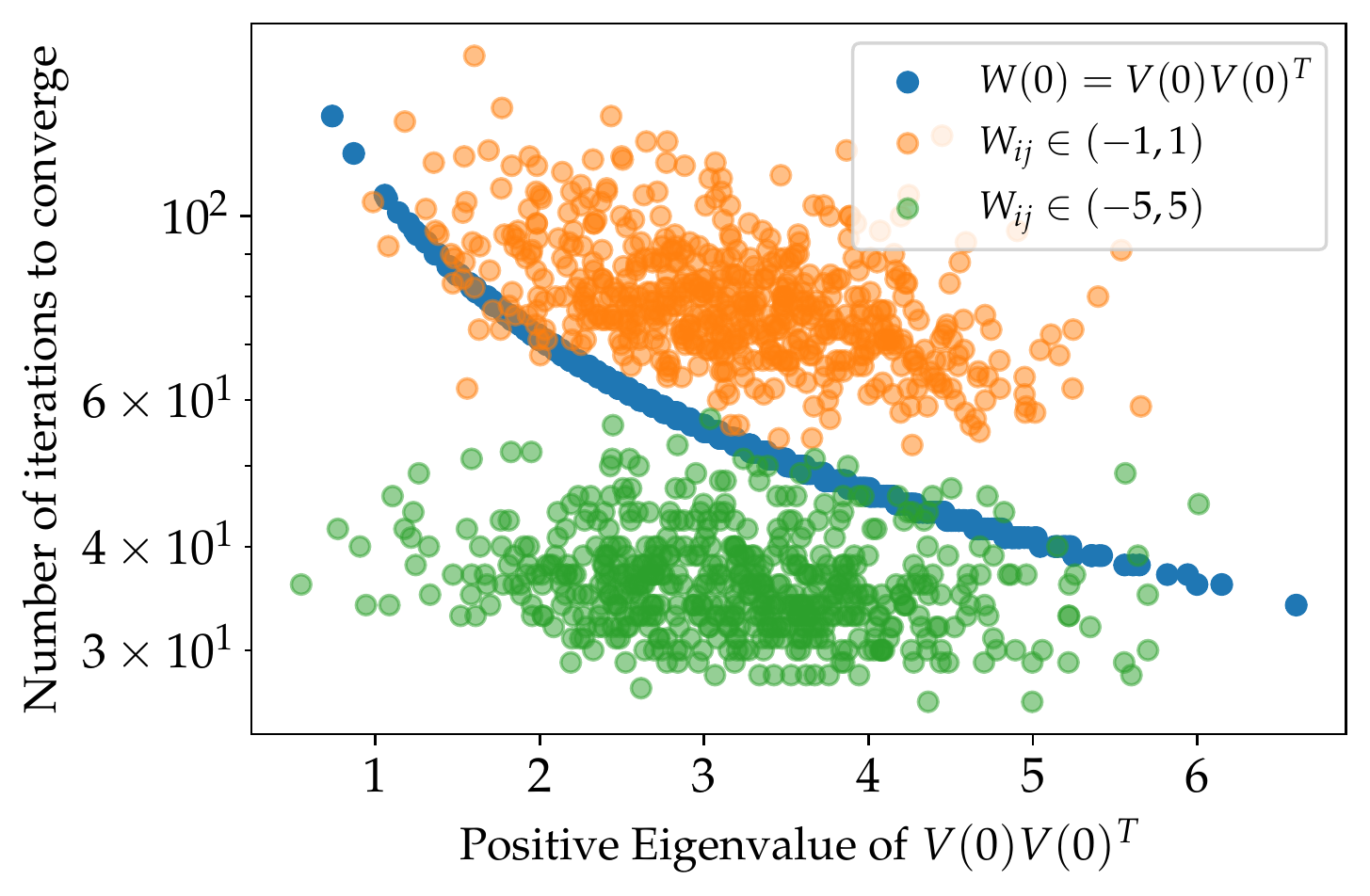}
\label{fig:convergence_3}
}

\subfigure[Erd\H{o}s-Renyi Model]{
\centering
\includegraphics[width=0.3\columnwidth]{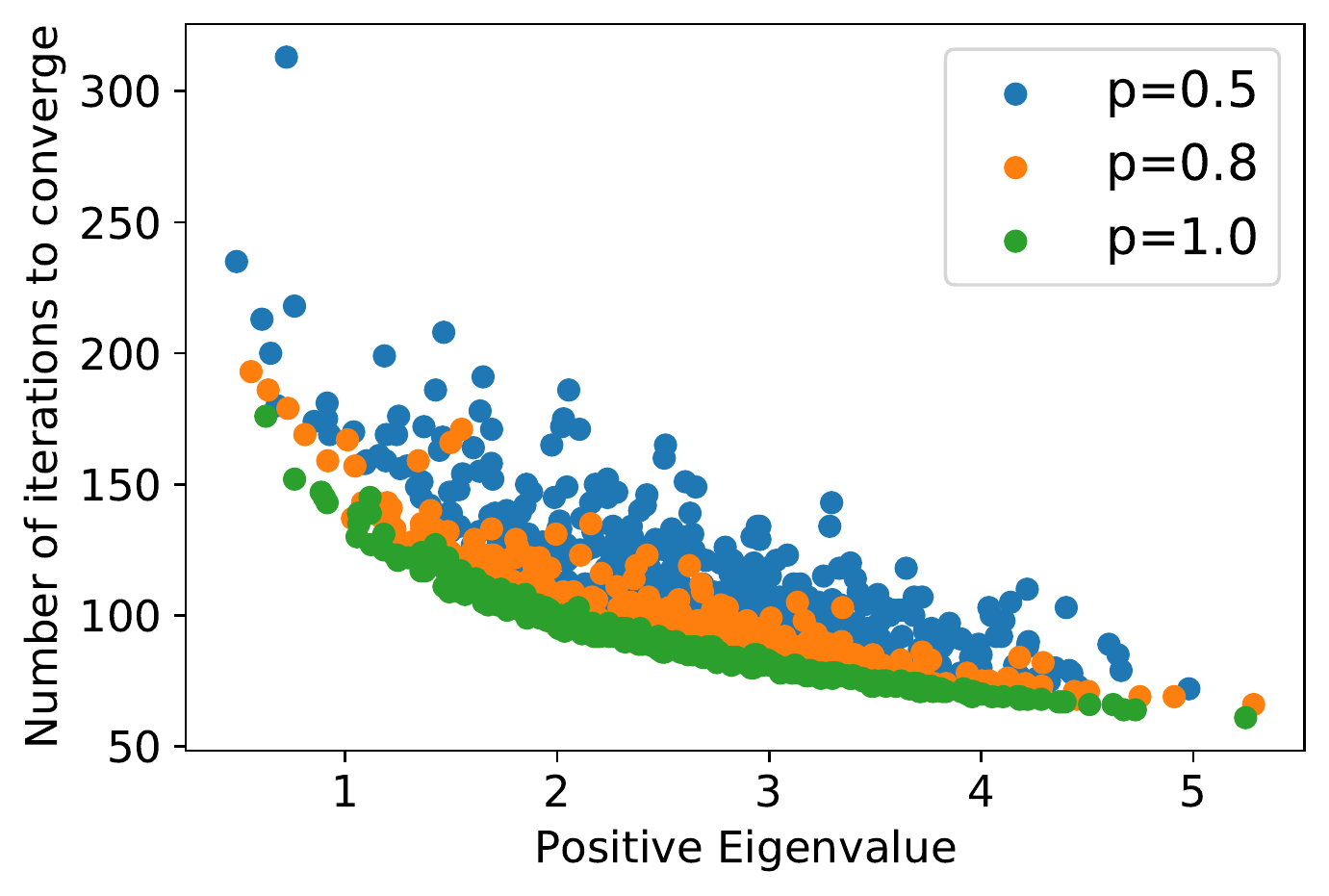}
\label{fig:convergence_4}
}
\subfigure[Watts-Strogatz Model ($p=0$)]{
\centering
\includegraphics[width=0.3\columnwidth]{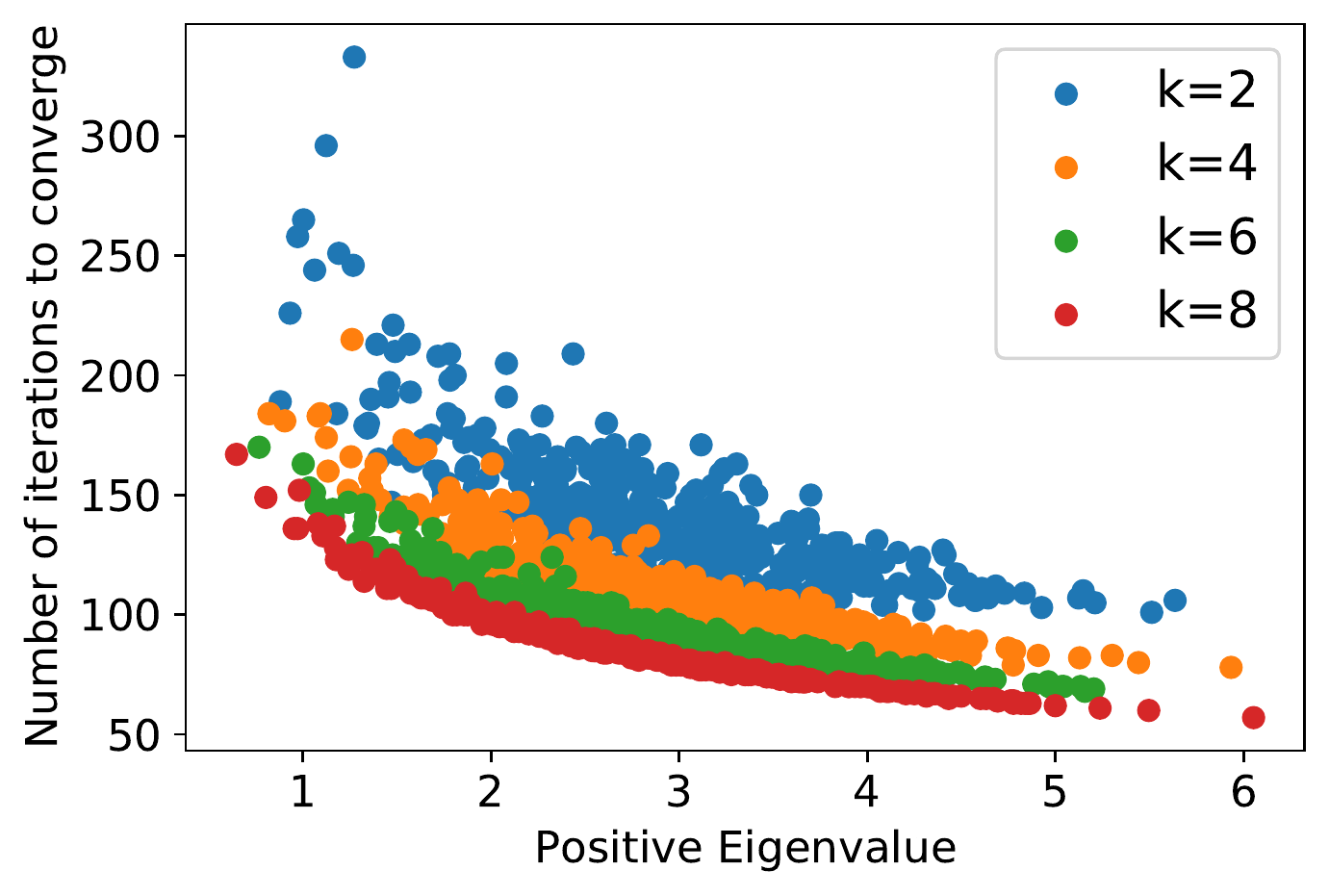}
\label{fig:convergence_5}
}
\subfigure[Watts-Strogatz Model ($p=0.5$)]{
\centering
\includegraphics[width=0.3\columnwidth]{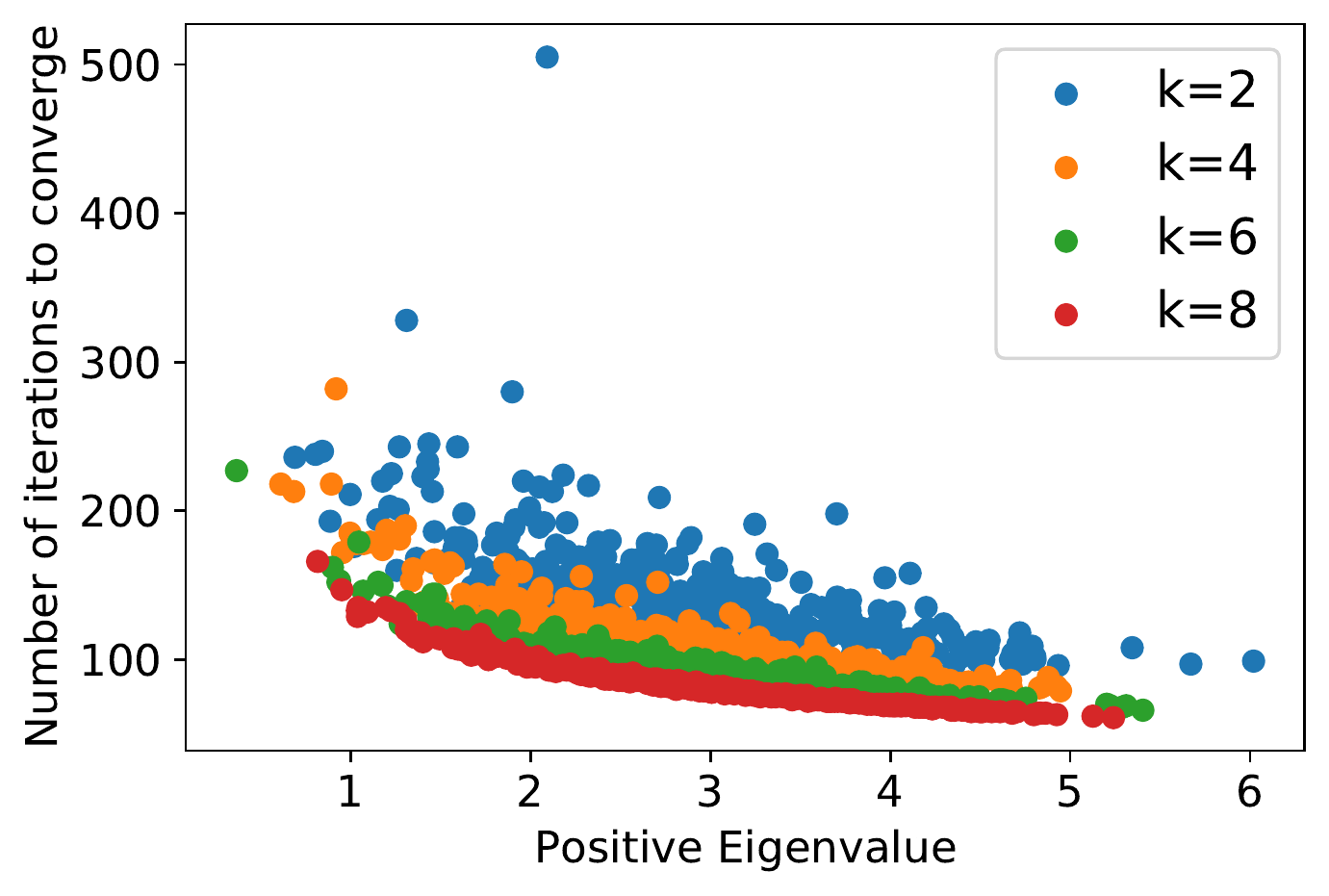}
\label{fig:convergence_6}
}
\caption{The number of iterations till network convergence. 
When the initial opinions are small in magnitude, the number of iterations to converge is large. When the graph becomes denser, the convergence rate increases.}
\label{fig:converge_rate}
\end{figure}

\smallskip\noindent\textbf{Magnitude of Initial Opinions.} Figure~\ref{fig:convergence_1} and ~\ref{fig:convergence_2} show two cases $W(0) = I$ and $W(0) = 0$ respectively. In both cases $BC=CB$. The initial opinions are randomly selected from $(-1, 1)$. We check the number of iterations until all entries in the opinion vector and weight matrix have absolute value larger than $10^{20}$. From the analysis in the previous section, the largest eigenvalue of $W(t)$ determines the convergence rate. Given that the initial weights are determined, we check the positive eigenvalue of $V(0)V(0)^T$. We can see that the number of iteration until convergence is inversely proportional to the magnitude of the positive eigenvalue of $V(0)V(0)^T$. The more extreme the initial opinions are, the faster the network reaches convergence. We also tested networks of different sizes, which seems to be generally oblivious to the convergence rate. 

\smallskip\noindent\textbf{$W(0)$ as a Random Matrix.} In Figure~\ref{fig:convergence_3}, we compare the case of $W(0) = V(0)V(0)^T$ with $W(0)$ being a random symmetric matrix ($W(0)$ and $C$ are generally not commutative). The orange dots show the number of iterations till convergence when entries in $W(0)$ are selected uniformly in $(-1, 1)$. The green dots show the number of iterations until convergence when entries in $W(0)$ are selected uniformly in $(-5, 5)$. There are a few observations. 1) When $W(0)$ take random values, it requires more iterations to convergence compared to the case when $W(0) = V(0)V(0)^T$. Specifically, during the evolution when $W(0) = V(0)V(0)^T$ both opinions and weights do not change signs. 2) When the initial weights take greater absolute values in general, the system converges faster. Again the more extreme the opinions/weights are, the faster the system reaches structural balance. 

\smallskip\noindent\textbf{On a General Graph.} We also test on graphs generated by social network models. In Figure~\ref{fig:convergence_4}, we show the convergence results on graphs generated by the Erd\H{o}s-Renyi Model $G(n, p)$, where $n$ is the number of nodes and $p$ is the probability of each pair of nodes connected by an edge. As $p$ increases from $0$ to $1$, the graph becomes denser and it requires fewer iterations to converge. 

In Figure~\ref{fig:convergence_5} and ~\ref{fig:convergence_6}, we tested on Watts Strogatz model~\cite{Watts1998-gi} $G(n, k, p)$, where $n$ is the number of nodes, $k$ is the number of neighbors of each nodes and $p$ is the rewiring probability. The network starts as a regular ring lattice, where each node connects to $k$ nearest neighbors on the ring. Each neighbor has a probability $p$ to be `rewired' to another non-neighbor node. Thus, with a larger value of $k$, there are more edges in the graph. With a larger rewiring probability $p$, there is more randomness in the graph. In these simulations, the initial weight matrices are set as a zero matrix, i.e., $W(0) = 0$.
Similarly, with the same number of edges in the graph, the relationship between the positive eigenvalue and the number of iterations follows the similar trend. When there are more edges in the graph, the convergence speed increases. In all cases, the network reaches structural balance.

\subsection{Emergence of Community Structure}
In this section, we ran experiments on real-world network data sets. We are particularly interested in the following question. Can a few planted seeds with opposite opinions influence the other nodes and drive the network into structural balance and polarization? Does the final state coincide with the community structure in the network? 

Our first experiment is based on a study by Zachary~\cite{zachary1977information} who witnessed the breakup of a karate club into two small clubs. As shown in Figure~\ref{fig:karate_1}, the networks captures $34$ members, documenting links between pairs of members who interacted outside the club. During the study, a conflict arose between the administrator (label $0$) and the instructor (label $33$), which led to the split. The red and green nodes represent the choice of each individual in the end. In this experiment, we assign the administrator and the instructor opposite opinions as $1$ and $-1$. The other members start with opinion $0$. Given that the links represent interaction and positive friendship between members, each edge is assigned a small positive value in the initial matrix $W(0)$. We then run our co-evolution dynamics till convergence. The edges with negative weight are removed. The graph is separated into two communities, shown in Figure~\ref{fig:karate_2}. It nearly predicted the same division as in the ground truth except for two members ($\#8$ and $\#19$) which are somewhat ambiguous.

The second experiment is based on the political blogs network. It is a directed network of hyperlinks between weblogs on US politics, recorded in 2005 by Adamic and Glance~\cite{adamic2005political}. There are $1,490$ nodes and $19,025$ directed edges in the graph. Each node has its political preference ($-1$ as liberal, $1$ as conservative) shown in Figure~\ref{fig:blog_1}. We randomly select $20\%$ nodes and assign initial opinions according to their ground truth values. All edges are assigned initial weights as a small positive value. When the graph reaches convergence, two big communities appear, as indicated by their final opinions and the sign of edges. 
Figure~\ref{fig:blog_2} shows the detected communities after negative edges are removed. 
Compared with the ground truth, the predicted opinions by our dynamical model has an accuracy of $97.21\%$, averaged by $200$ simulation runs. If only $3\%$ nodes are assigned ground truth opinions in the initial state, the prediction accuracy for the final opinions of all nodes, on average, is as high as $82.12\%$. 

Our dynamic model, as shown by these experiments, explains why community structures appear. It can also be understood as an algorithm for label propagation or node classification. Compared with other methods for the same task~\cite{Xie2013-hz, Jokar2019-km,Tang2016-vp, Bhagat2011-rt} that generally use data-driven machine learning approaches, our dynamic model has better transparency and interpretability. 